\documentclass[
aps,
prd,
10pt,
twocolumn,
nofootinbib,
preprintnumbers,
superscriptaddress,
showpacs]{revtex4-2}

\usepackage{comment}
\usepackage{booktabs,makecell}
\usepackage{graphicx}
\usepackage{amsmath,amssymb}
\usepackage{amsfonts}
\usepackage{xspace} 
\usepackage[usenames]{color}
\usepackage{dcolumn}
\usepackage{bm}
\usepackage{mathrsfs}
\usepackage[colorlinks=true,citecolor=blue]{hyperref}
\usepackage[all]{hypcap} 
\usepackage[utf8]{inputenc} 
\usepackage{slashed}
\usepackage{multirow}
\usepackage{rotating}
\definecolor{orange}{rgb}{1,0.5,0}
\usepackage{tabularx}
\usepackage{siunitx}
\usepackage[normalem]{ulem}
\usepackage[dvipsnames]{xcolor}

\def\rrBM{\ifmmode r^{\text{BM}}_{\text{ratio}} \else {$r^{\text{BM}}_{\text{ratio}}$}\fi}
\def\rrDM{\ifmmode r^{\text{DM}}_{\text{ratio}} \else {$r^{\text{DM}}_{\text{ratio}}$}\fi}

\def\Msun{{\text{\,M}_\odot}}

\graphicspath{{./}{Images/}}
\begin{document}

\title{Hybrid Stars with Post-Merger Rotation Profiles}

\author{Kalin V. Staykov}\email{kstaykov@phys.uni-sofia.bg}
\affiliation{Department of Theoretical Physics, Faculty of Physics,
Sofia University ``St. Kliment Ohridski", Sofia 1164, Bulgaria}

\author{Violetta Sagun}\email{v.sagun@soton.ac.uk}
\affiliation{Mathematical Sciences and STAG Research Centre, University of Southampton, Southampton SO17 1BJ, United Kingdom}

\author{Lorenzo Cipriani}\email{lorenzo.cipriani@graduate.univaq.it}
\affiliation{Dipartimento di Scienze Fisiche e Chimiche, Università dell’Aquila, via Vetoio, I-67100, L’Aquila, Italy}
\affiliation{INFN, Laboratori Nazionali del Gran Sasso, I-67100 Assergi (AQ), Italy}

\author{Daniela D. Doneva}\email{daniela.doneva@uni-tuebingen.de}
\affiliation{Departamento de Astronom\'ia y Astrof\'isica, Universitat de Val\`encia,
Avinguda Vicent Andrés Estellés 19, 46100, Burjassot (Val\`encia), Spain}
\affiliation{Theoretical Astrophysics, Eberhard Karls University of T\"ubingen, 72076 T\"ubingen, Germany}

\author{Stoytcho S. Yazadjiev}\email{yazad@phys.uni-sofia.bg}
\affiliation{Department of Theoretical Physics, Sofia University ``St. Kliment Ohridski", Sofia 1164, Bulgaria}
\affiliation{Institute of Mathematics and Informatics, Bulgarian Academy of Sciences, Acad. G. Bonchev St. 8, Sofia 1113, Bulgaria}

\begin{abstract}
We study the effect of differential rotation on hybrid stars with the first-order deconfinement phase transition from hadronic to color superconducting quark matter. The differential rotation is introduced within a realistic, four-parameter phenomenological rotation law, in which the maximum angular velocity of the rotating configuration is shifted away from the center. We focus on two classes of differentially rotating solutions, namely quasi-toroidal (type C) and quasi-spherical (type A), and study the changes in the star global properties and angular velocity profiles due to the presence of a phase transition. We show that when increasing the angular momentum $J$ the turning points of the $J=const$ sequences shift towards lower energy densities, shrinking considerably the region where differentially rotating neutron stars with phase transitions exists. Interestingly, for both type A and type C solutions, the angular velocity profile is continuous throughout the star despite the discontinuity in the energy density. Moreover, we show that for the differential rotation law we study, at the crossing points where the mass-radius curves for different equations of state intersect, the rotational profiles of the solutions are very close despite large differences in the energy density profiles. This reveals a possible degeneracy between the post-merger remnant
properties for models with and without phase transitions, emphasizing the need for complementary multi-messenger observables to distinguish between them.
\end{abstract}

\maketitle

\section{Introduction}
\label{sec:intro}
Neutron stars (NSs) are among the most fascinating compact objects observed in the Universe, offering an unprecedented opportunity to indirectly probe the properties of matter at extreme densities and conditions that cannot be achieved on Earth~\cite{Baym:2017whm}. The plethora of observational manifestations, both in the electromagnetic and gravitational wave (GW) spectrum, provides a rich phenomenology to test different hypotheses~\cite{Ozel:2016oaf,Baiotti:2016qnr}. 

An important characteristic of NSs is that they are always rotating, with rotational frequencies spanning a wide range, reaching up to 716 Hz for the two fastest observed pulsars~\cite {Hessels:2006ze,Jaisawal:2024lps}. At such high rotational rates, the star's deformation from spherical symmetry becomes non-negligible, and the slow-rotation approximation can no longer be safely applied~\cite{Paschalidis:2016vmz}. Due to the high compactness of NSs, they can theoretically reach even higher rotational rates. During their evolution, though, the slowdown due to effects such as rotation instabilities~\cite{Andersson:2000mf} and magnetic dipole radiation will effectively set an upper limit on the rotational rate for old NSs (see e.g.~\cite{Haskell:2018nlh,Chakrabarty:2003kt}).

The situation is very different in the first few tens of milliseconds following the merger of two NSs. In that case, the angular momentum of the newly formed compact object is very high, and it can be supported against collapse due to differential rotation, thus forming a supramassive or hypermassive NS~\cite{Duez:2004nf,Baiotti:2016qnr}. For GW170817, multimessenger modelling suggests that the merger remnant survived for a timescale of order $\sim 0.1$–$1$ s as a differentially rotating hypermassive NS before collapsing to a black hole~\cite{LIGOScientific:2018cki,Gill:2019bvq}. Even though the post-merger phase is a highly dynamical process, the newly formed object can be accurately approximated by stationary NSs with a properly chosen differential rotation law~\cite{Uryu:2017obi,Cassing:2024dxp,Passamonti:2020yvh} and possibly a nontrivial temperature profile~\cite{Camelio:2019rsz,Camelio:2020mdi,Iosif:2020iho,Iosif:2021-09312}. Even though such treatment is, of course, less accurate compared to results from numerical relativity simulations, its power lies in the much faster generation of models and thus the ability to explore the parameter space in detail. Moreover, stationary models have already proven useful in interpreting the GW post-merger spectrum, in estimating threshold masses for prompt collapse, and constructing empirical equation of state (EoS) independent relations~\cite{Paschalidis:2016vmz,Bozzola:2017qbu,Weih:2017mcw, Iosif:2020iho,Ciolfi2021, Rosati2021,Jaraba:2026nmq}.

Apart from the difficulty of properly modeling a post-merger remnant with a stationary solution and the associated uncertainty in the adopted differential rotation law, another level of complexity is the underlying EoS. Quantum chromodynamics predicts that dense matter transitions from hadronic to quark matter at sufficiently high densities; however, the onset density of this phase transition remains uncertain~\cite{Annala:2019puf}. Therefore, it is not yet clear whether the cores of isolated NSs attain densities high enough to trigger the formation of quark matter, whether the quark matter can be created during a binary NS (BNS) merger when the density and temperature rise, or whether the transition occurs only at densities beyond those realized in NSs~\cite{Alford:2006vz,Weissenborn:2011qu,Chamel:2012ea,Ivanytskyi:2019ojt,Somasundaram:2021clp,Gartlein:2023vif,Christian:2023hez,Gartlein:2025zhd}. The deconfinement phase transition between hadronic and quark matter can have significant astrophysical implications, particularly during NS mergers~\cite{Bauswein:2018bma,Han:2018mtj,Weih:2019xvw,Most:2018eaw,Prakash:2021wpz,Ujevic:2022nkr,Blacker:2024tet,Hammond:2025kki}. It is natural to explore in a complementary, computationally less demanding way, whether such a phase transition will influence the structure and properties of the post-merger remnant, especially taking into account that in the early post-merger phases, one can observe an off-center maximum of the mass. 

Uniformly rotating stars with deconfinement phase transitions have already shown interesting new characteristics~\cite{Bozzola:2019tit,Gartlein:2024cbj} while differential rotation with the so-called $J$-constant differential rotation law and piecewise polytropic EoS, including universal relation construction, was considered in~\cite{Bozzola:2019tit}. Our aim is to perform a systematic study of differentially rotating hybrid stars implementing a realistic post-merger differential rotation law by~\citet{Uryu:2017obi} and realistic tabulated EoS. Compared to previous studies, our approach allows for an off-center maximum in the angular velocity, which may be observable in the first milliseconds following the formation of a merger remnant. We consider six hybrid configurations with different quark onset densities and quark matter properties, enabling a systematic investigation of the impact of the phase transition on the angular velocity profile. Furthermore, we examine how the angular velocity differs between various pairs of models at the crossing points at which the mass–radius curves for different EoSs intersect~\cite{Gartlein:2023vif,Cierniak:2020eyh}.

The paper is organized as follows. In Section~\ref{sec:EoS} we describe the hadronic and quark EoSs as well as the hybrid star construction, while in Section~\ref{sec:diff} we discuss the theoretical background for the differential rotation. In Section~\ref{sec:results} we present the results of the numerical calculation of NS models, particularly for the quasi-toroidal~\ref{sec:results_tor}, quasi-spherical~\ref{sec:results_sph} solutions, and we compare the solutions in the mass-radius intersection points in~\ref{sec:special_points}. Finally, Section~\ref{sec:conclusions} summarizes our findings.

We employ several types of units. Thermodynamic quantities are presented in cgs units, baryonic density in units of the nuclear saturation density ($n_{sat} = 2.48\times10^{14} \text{ g/cm}^3$), while the star radius is shown in km. The mass is in units of the Solar mass. 

\section{Hybrid stars}
\label{sec:EoS}

\subsection{Hadronic matter EoS}
\label{subsec:NMEoS}

The hadronic phase is modeled within the relativistic density functional DD2npY-T EoS~\cite{Shahrbaf:2022upc}, which reproduces the key nuclear matter ground state properties, satisfies the observational constraints on the maximum NS mass~\cite{Fonseca:2021wxt,Romani:2021xmb}, tidal deformability inferred from the GW170817 BNS merger~\cite{LIGOScientific:2018cki}, and the NICER measurements~\cite{Miller:2019cac,Vinciguerra:2023qxq,Salmi:2024aum,Dittmann:2024mbo,Choudhury:2024xbk,Salmi:2024bss,Mauviard:2025dmd}. The DD2npY-T EoS was chosen as a realistic EoS with both nucleonic and hyperonic degrees of freedom with a maximum mass reaching 2.154$M_{\odot}$. Below the nuclear saturation density, it is matched, within a unified framework, to a crust EoS~\cite{Typel:2018wmm}. 

\subsection{Quark matter EoS}

\begin{figure}[ht]
	\includegraphics[width=0.45\textwidth]{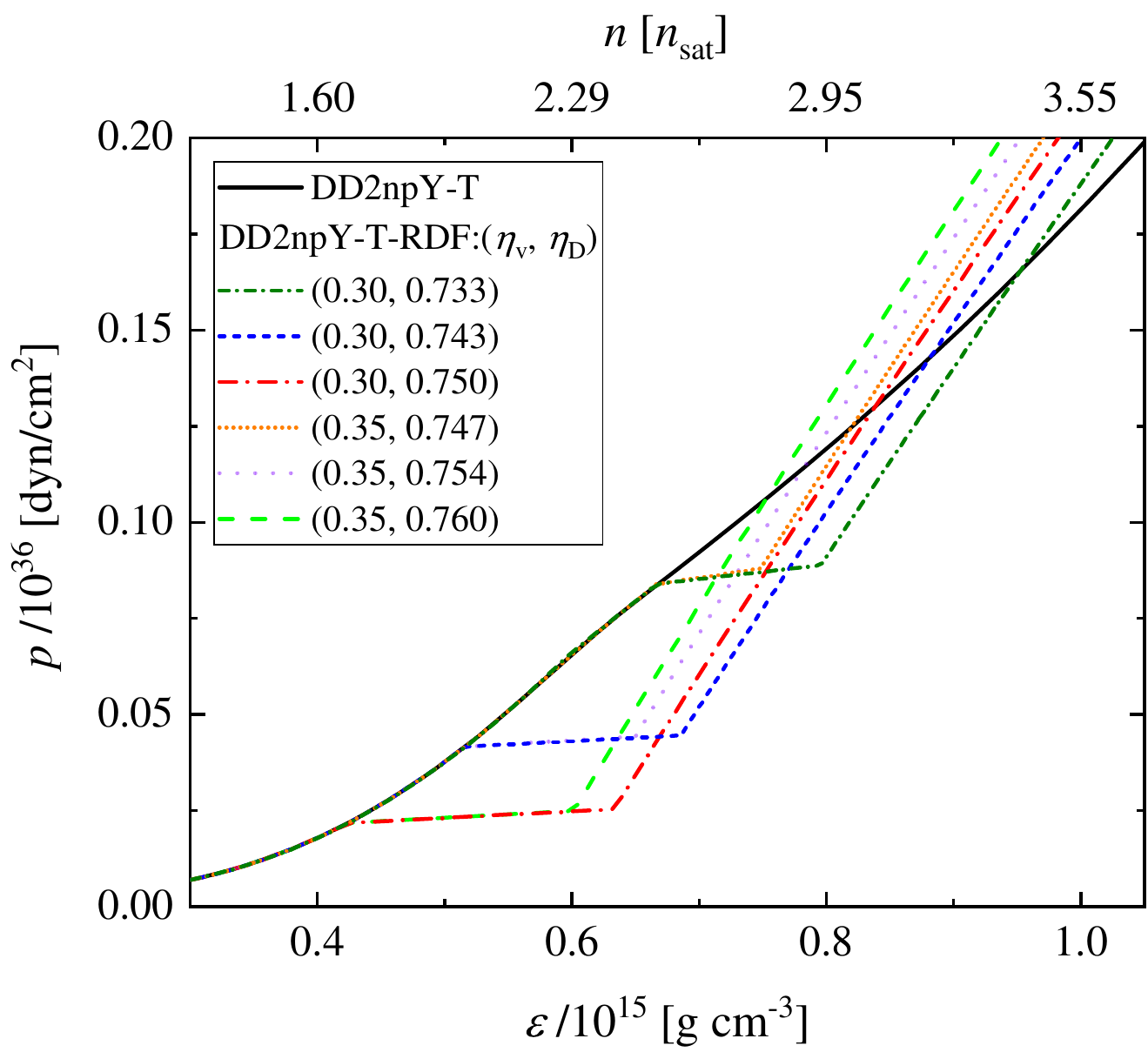}
	\caption{Pressure as a function of the energy density (bottom x-axis) and baryonic density (top x-axis) for all EoS considered with different combinations of $\eta_V$ and $\eta_D$.}
	\label{fig:EoSs}
\end{figure}

Quark matter is described within the relativistic density functional (RDF) approach with the medium-dependent scalar and pseudoscalar couplings~\cite{Ivanytskyi:2022oxv,Ivanytskyi:2022bjc}. The vector and diquark couplings acquire a medium dependence that is adjusted to ensure that quark matter approaches the conformal limit at high densities~\cite{Ivanytskyi:2022bjc}. The model incorporates quark pairing in the two-flavor color-superconducting (2SC) phase. Its two key parameters are the dimensionless vector coupling $\eta_V$  and the diquark pairing strength $\eta_D$. Increasing $\eta_V$ strengthens the repulsive interaction between quarks, thereby stiffening the EoS, while $\eta_D$ governs the onset of quark matter. Following~\cite{Gartlein:2023vif,Gartlein:2025}, we chose two different values for the vector coupling $\eta_V$=0.30 and $\eta_V$=0.35 and consider several $\eta_D$ values to cover a wide range of quark matter onset densities. The full set of parameters for the considered hybrid EoS are presented in Table~\ref{tbl:Param_table}. In Fig.~\ref{fig:EoSs}, the pressure is shown as a function of the energy density (bottom x-axis) and as a function of the baryon density (top x-axis) for all configurations. Only the phase transition region is displayed.
\begingroup
\setlength{\tabcolsep}{10pt} 
\renewcommand{\arraystretch}{1.5} 
\begin{table}[!h]
\begin{center}
\footnotesize
\caption{The parameters of the considered hybrid configurations: the values of the vector and diquark couplings, baryonic density of the deconfinement phase transition onset to color superconducting quark matter, and the maximum mass of the static configuration.}
{\begin{tabular}{@{}ccccc@{}}
 \hline
  $\eta_V$ \quad & $\eta_D$ & $n_{\textrm onset}$ [$n_{sat}$] & $M_{onset}$ [$M_{\odot}$]  & $M_{max}$ [$M_{\odot}$] \\
 \hline
 0.30   & 0.733  &  2.517 & 0.711 & 2.177\\
 0.30   & 0.743  &  2.009 & 1.067 & 2.179\\
 0.30   & 0.750  &  1.684 & 1.534 & 2.292\\
 0.35   & 0.747  &  2.517 & 0.711 & 2.274\\
 0.35   & 0.754  &  2.009 & 1.067 & 2.275\\
 0.35   & 0.760  &  1.684 & 1.534 & 2.280\\
 \hline
 \multicolumn{2}{c}{DD2npY-T} & NA & NA & 2.154\\
 \hline
\end{tabular}}
\label{tbl:Param_table} 
\end{center}
\end{table}

\subsection{Static hybrid configurations}

Hybrid stars encompass both hadronic and quark phases, with a deconfinement phase transition in between. The onset of the phase transition is determined using the Maxwell construction, resulting in a sharp interface between the phases corresponding to the first-order phase transition. The Maxwell construction is based on fulfilling the condition of mechanical stability and baryon chemical equilibrium at the phase boundary, and is defined as the intersection of pressures as a function of the chemical potential for hadronic and quark phases
\begin{equation}
    p^{hadron}(\mu_B) = p^{quark}(\mu_B),
\end{equation}
where the baryonic chemical potential $\mu_B$ corresponds to the onset chemical potential $\mu_B^{onset}$. At the same time, the Maxwell construction leads to the discontinuous behaviour of the baryonic density 
\begin{equation}
    n_B^{hadron}(\mu_B^{onset}) \ne n_B^{quark}(\mu_B^{onset}).
\end{equation}
The jump in the baryonic density defines the width of the mixed phase region, or the width of the horizontal plateau. 
\begin{figure*}[ht]
 \includegraphics[width=0.95\textwidth]{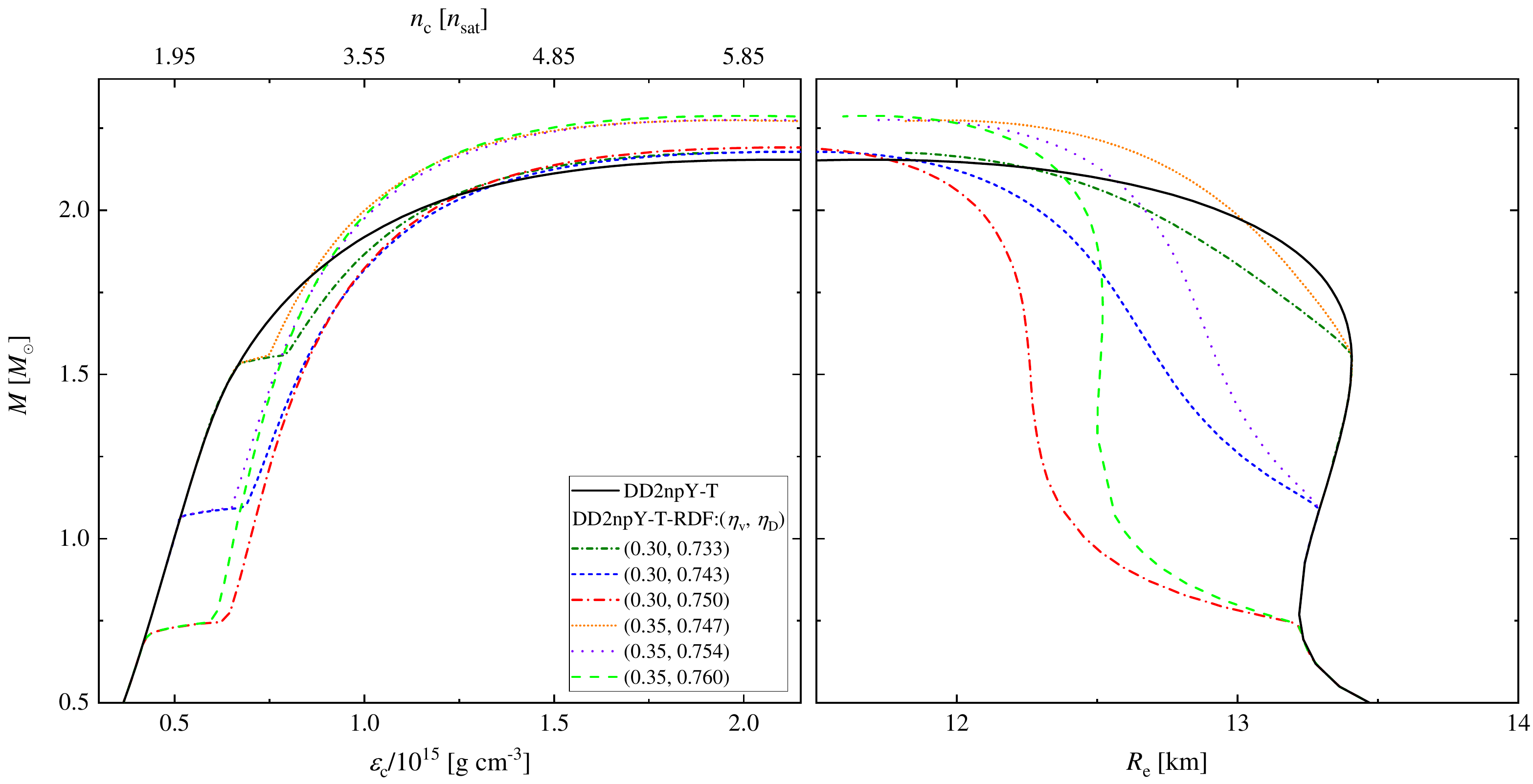}
	\caption{\textit{Left panel:} the gravitational mass of static configurations as a function of both the central energy density in $\text{g/cm}^3$ (bottom $x$-axis) and central baryonic density (upper $x$-axis), expressed in units of the nuclear saturation density. \textit{Right panel:} the mass-radius relations for the considered EoSs.}
	\label{fig:M_rho_stat}
\end{figure*}
This discontinuous behavior of the baryonic density and energy density poses a numerical challenge for many codes. The numerical integration of the field equations involves interpolations and numerical derivatives of the hydrodynamical quantities, which prevents the code from handling the discontinuities in the EoS associated with the phase transition. Therefore, to resolve this, a slight smoothing is introduced. As we will see in the remainder of the paper, the smoothing will manifest as a finite area in the star between the pure hadronic phase and color-superconducting quark matter, rather than an infinitely thin transition. 

An alternative approach to overcoming the numerical challenge is to consider the Glendenning construction or pasta phase, in which not only the pressure varies continuously but also the baryonic density~\cite{Glendenning:1992vb}. Compared to the phase transition construction adopted in this work, the Glendenning construction yields a much smoother transition and, consequently, a more extended mixed phase.

To model the structure of static spherically symmetric stars, we solve the Tolman-Oppenheimer-Volkoff equations (TOV)~\cite{Oppenheimer:1939ne,Tolman:1939jz}. The left panel of Fig.~\ref{fig:M_rho_stat} presents the gravitational mass as a function of the central energy density (bottom x-axis) and baryonic density (upper x-axis), while the right panel presents the gravitational mass as a function of the radius of the star. In both panels, the pure baryonic DD2npY-T EoS (solid black curves) and all hybrid configurations from Table~\ref{tbl:Param_table} (colorful curves) are presented. The selected parameter set represents six distinct hybrid star configurations, corresponding to early, intermediate, and late onsets of deconfinement, thereby adopting a more agnostic perspective on the properties of the hadron–quark phase transition. For each of the three deconfinement onset densities, we consider two configurations with different stiffness, which allows us to study the effect of quark matter stiffness on the results (e.g. the red and green, blue and purple, or orange and dark green curves in Fig.~\ref{fig:M_rho_stat}). In the left panel of Fig.~\ref{fig:M_rho_stat}, the jumps in energy density (horizontal plateaus) mark the mixed hadronic-quark phase.

The intersection of the considered mass–radius curves gives rise to the so-called {\it crossing points} (CPs). In the static case, these configurations are referred to in the literature as {\it special points}~\cite{Cierniak:2020eyh,Cierniak:2021knt,Cierniak:2021vlf}. In Ref.~\citep{Gartlein:2023vif}, a family of special points was used to establish an empirical relation between the mass at the special point, the maximum mass of the corresponding mass–radius curve, and the onset mass of quark deconfinement. Therefore, this provides a way to relate the microscopic parameters of the EoS, such as the coupling constants, to macroscopic stellar observables, in particular the NS mass. 
To distinguish the intersection points for differentially rotating vs. static configurations, we refer to them as CPs. The question of how special points are modified by increasing angular velocity and changing the rotation law is an interesting one that warrants further investigation. We leave it for future studies. 

\section{Differential rotation}
\label{sec:diff}

In the present paper, we consider stationary and axisymmetric matter described by the following line element
\begin{eqnarray} \label{eq:metric}
ds^2 &= &-e^{\gamma+\sigma} dt^2 + e^{\gamma-\sigma} r^2
\sin^2\theta (d\phi - \omega dt)^2 + \nonumber\\ && e^{2\alpha}(dr^2 + r^2
d\theta^2),
\end{eqnarray}
where all metric functions depend on $r$ and $\theta$ only. To solve the field equations, we use an extended version of the ${\tt RNS}$~\cite{1995ApJ...444..306S} code, which is based on the modified~\cite{1994ApJ...422..227C} Komatsu-Eriguchi-Hachisu (KEH) scheme~\cite{Komatsu:1989ikr, 1989MNRAS.239..153K} for finding the rotating configurations (other codes modeling rapidly rotating NSs include XNS~\cite{Pili:2016hqo,Franceschetti:2022ypc}, LORENE~\cite{lorene}, COCAL~\cite{Uryu:2017obi} and the recently developed extension of FUCA~\cite{Tootle:2026ebk}). Differential rotation is implemented according to the procedure described in~\cite{Iosif:2020iho, Iosif:2021-09312,Staykov:2023ose}. 

According to numerical relativity simulations, a general property of the post-merger remnants in the first milliseconds after their formation is that the angular momentum peaks away from the center~\cite{Baiotti:2016qnr}. One of the simplest realistic rotational laws that allows for such a structure is the phenomenological four-parameter law by~\citet{Uryu:2017obi}, which has already been successfully implemented in general relativity (GR)~\cite{Iosif:2020iho, Iosif:2021-09312}, modified theories of gravity~\cite{Staykov:2023ose,Lam:2025jsk}, dark matter admixed NSs~\cite{Cipriani:2025wem}, and in a more involved study comparing realistic differential rotation laws with merger remnants from full GR simulations~\cite{Cassing:2024dxp}. 

The four-parameter law\footnote{Note that one should understand the term differential rotation law in a broader sense commonly used in the literature. In particular, eq. \eqref{Eq:DiffRotLaw} does not directly prescribe the angular velocity as a function of position inside the star $\Omega(r,\theta)$. Instead, it specifies a functional relation $\Omega=\Omega(F)$ between the angular velocity and the gravitationally redshifted angular momentum, which enters in the equation for hydrostationary equilibrium.} adopted in the present study reads
\begin{equation} \label{Eq:DiffRotLaw}
    \Omega = \Omega_c \frac{1 + \left(\frac{F}{B^2\Omega_c}\right)^p}{1 + \left(\frac{F}{A^2\Omega_c}\right)^{p+q}},
\end{equation}
where $F = u^t u_{\phi}$ is the gravitationally redshifted angular momentum per unit rest mass and enthalpy, and $p, q, A,$ and $B$ are parameters. The free parameters $p$ and $q$ are fixed to $p=1$ and $q=3$, which allows the equation for hydrostationary equilibrium to be cast in an explicit algebraic form required for the procedure implemented in the ${\tt RNS}$ code. For the parameters $A$ and $B$, we follow ~\cite{Iosif:2020iho} (and the references therein) and instead of fixing them, we introduce new parameters defined by the ratios of the maximum angular velocity and the equatorial angular velocity with respect to the angular velocity in the center (all defined with respect to an observer at infinity)
\begin{eqnarray} 
\label{eq:l1}
&&\lambda_1 = \frac{\Omega_{max}}{\Omega_c}, \\
\label{eq:l2}
&&\lambda_2 = \frac{\Omega_{e}}{\Omega_c}.
\end{eqnarray}
Given those ratios, the parameters A and B can be obtained by solving the equations derived by substituting the explicit form of the angular velocity \eqref{Eq:DiffRotLaw} into Eqs. \eqref{eq:l1}-\eqref{eq:l2}. This choice of reparameterization has been widely used since the original paper of~\citet{Uryu:2017obi} in a number of studies~\cite{Zhou:2019hyy,Iosif:2020iho,Iosif:2021-09312,Staykov:2023ose,Cassing:2024dxp}. Since this rotation law is used mainly in the context of simulating post-merger remnants, an advantage of using the parameters $\lambda_1$ and $\lambda_2$ instead of $A$ and $B$ is that the ratios \eqref{eq:l1} and \eqref{eq:l2} can be extracted from BNS merger simulations and more specifically, from the angular velocity profiles of the post-merger remnant. Hence, more realistic differentially rotating solutions can be achieved. An additional benefit is that $\lambda_2$ directly controls the type of solution, distinguishing between the quasi-spherical (type A) and quasi-toroidal (type C) configurations. For a more detailed discussion of how $\lambda_1$ and $\lambda_2$ affect the shape of the rotational profile, we refer the reader to Section 3.1 of Ref.~\cite{Cassing:2024dxp}.

Let us briefly explain the numerical procedure for computing a stationary neutron star solution. The KEH scheme is based on an integral representation of the field equations using Green functions and the use of the first integral of the relativistic Euler equation for hydrostationary equilibrium. The solutions are parametrized by the central/maximal energy density and a parameter which fixes the rate of rotation -- in our case this is the angular momentum. For the differential rotation the parameters $\lambda_1$ and $\lambda_2$ should be provided at the start as well. The iterations for constructing a solution with those fixed parameters start with an initial guess for the metric functions, the energy density, and the angular velocity distributions. The constants $A$ and $B$ in Eq.~\eqref{Eq:DiffRotLaw} are updated at the start of each iteration. During the iterative procedure, the equation for hydrostationary equilibrium and the field equations are used to update the matter distribution and the metric function until convergence is achieved. In the way the KEH method is designed and implemented in the ${\tt RNS}$ code, the differential rotation law is a free function chosen in such a way, that the first integral can be cast in an analytic form. For the rotational law \eqref{Eq:DiffRotLaw} this is possible only when $(p,q) = (1,3)$. This clearly imposes restrictions, but it has been shown to produce realistic post-merger differential rotation profiles \cite{Iosif:2021aum}. A thorough explanation of the KEH scheme with the implemented differential rotation law \eqref{Eq:DiffRotLaw} is given in \cite{Iosif:2020iho}.

In the present study, we follow the classification of the solutions introduced in~\cite{Ansorg:2008pk}. In this scheme, the type A solutions remain quasi-spherical as the rotation rate increases: the maximum energy density remains at the center as the ratio of polar to equatorial radius decreases. The type C solutions, on the other hand, evolve from quasi-spherical to quasi-toroidal with increasing rotation, with the maximum energy density shifting away from the center.

It is important to mention the limitations of the ${\tt RNS}$ code. The KEH procedure implemented in the code has a limited applicability for finding the different types of differentially rotating solutions. Of all types reported in~\cite{Ansorg:2008pk}, only types A and C can be reliably found with the ${\tt RNS}$ code. On the other hand, with the increase of the angular momentum $J$, the sequences of constant $J$ get progressively shorter as the code fails to converge to a unique solution. This is because, for a given differential rotation law, the numerical procedure does not discriminate between different types of solutions that share the same ratio of polar to equatorial radius, $r_p/r_e$, and the same central or maximum energy density. These limitations are known and discussed in~\cite{Iosif:2021-09312, Staykov:2023ose}. In addition, for the quasi-spherical solutions, there are additional convergence problems which depend on the pair $(\lambda_1,\lambda_2)$ as discussed in~\cite{Iosif:2021-09312}. As we will show later, this results in a significant reduction in the maximum angular momentum that can be reliably explored for quasi-spherical configurations.

GR hydrodynamic simulations~\cite{Paschalidis:2015mla,Zhang:2017fsy} indicate that, during the intermediate post-merger phase, the remnant frequently develops a quasi-toroidal differentially rotating configuration, which subsequently evolves toward a more quasi-spherical shape at later times. The evolution of the remnant can be followed in Fig. 7 and Fig. A1 in~\cite{Cassing:2024dxp}. The change in the angular velocity profile corresponds to the change in the ratio $\lambda_2 = {\Omega_e}/{\Omega_c}$ that determines the shape of the remnant. For the models presented in~\cite{Cassing:2024dxp}, at 4 to 6 ms of the post-merger phase (depending on the EoS) the rotational profile corresponds to roughly $\lambda_2 \sim 0.5$ (according to simulations in GR), which translates to a quasi-toroidal configuration. At later times, 14 to 22 ms (again, depending on the EoS), $\lambda_2 \sim 1$, therefore, the remnant evolves to be quasi-spherical.

\section{Results}
\label{sec:results}

The main objective of the present study is to examine the effects of differential rotation on hybrid stars, described by Uryu's differential rotation law~\eqref{Eq:DiffRotLaw}, and exhibiting the deconfinement phase transition from hadronic to color superconducting quark matter within a realistic EoS. The resolution employed for constructing the sequences of equilibrium solutions is $601 \times 301$ in radial and angular directions, respectively, while for the presented individual models, we increase it to $801 \times 401$ for smoother contour plots of the constant energy density values. 

As discussed above, the pure hadronic solutions are modelled with the DD2npY-T EoS, while the hybrid configurations contain a color superconducting quark matter cores with the first-order phase transition from hadronic matter. In the text we mostly present three representative parametrisations for the DD2npY-T-RDF EoS from the six listed in Table~\ref{tbl:Param_table} and Fig.~\ref{fig:EoSs}. We explored all six hybrid parameterizations and chose the representative ones so that they account for different quark matter stiffness and quark onset density, namely, we chose $(\eta_V = 0.35,\eta_D = 0.760)$ which is stiffer and the onset is at $0.7 \Msun$, and two softer ones with $(\eta_V = 0.30,\eta_D = 0.743)$ with onset at $1 \Msun$ and $(\eta_V = 0.30,\eta_D = 0.733)$ with onset at $1.5 \Msun$. As it turns out, for such values of the parameters, phase transition can be observed for both type A and type C solutions and for relatively high angular momenta. In addition, the same EoS sets were also used in previous studies of uniformly rotating hybrid stars~\cite{Gartlein:2024cbj}, which facilitates comparison.
\begin{figure}[ht]
   	\includegraphics[width=0.45\textwidth]{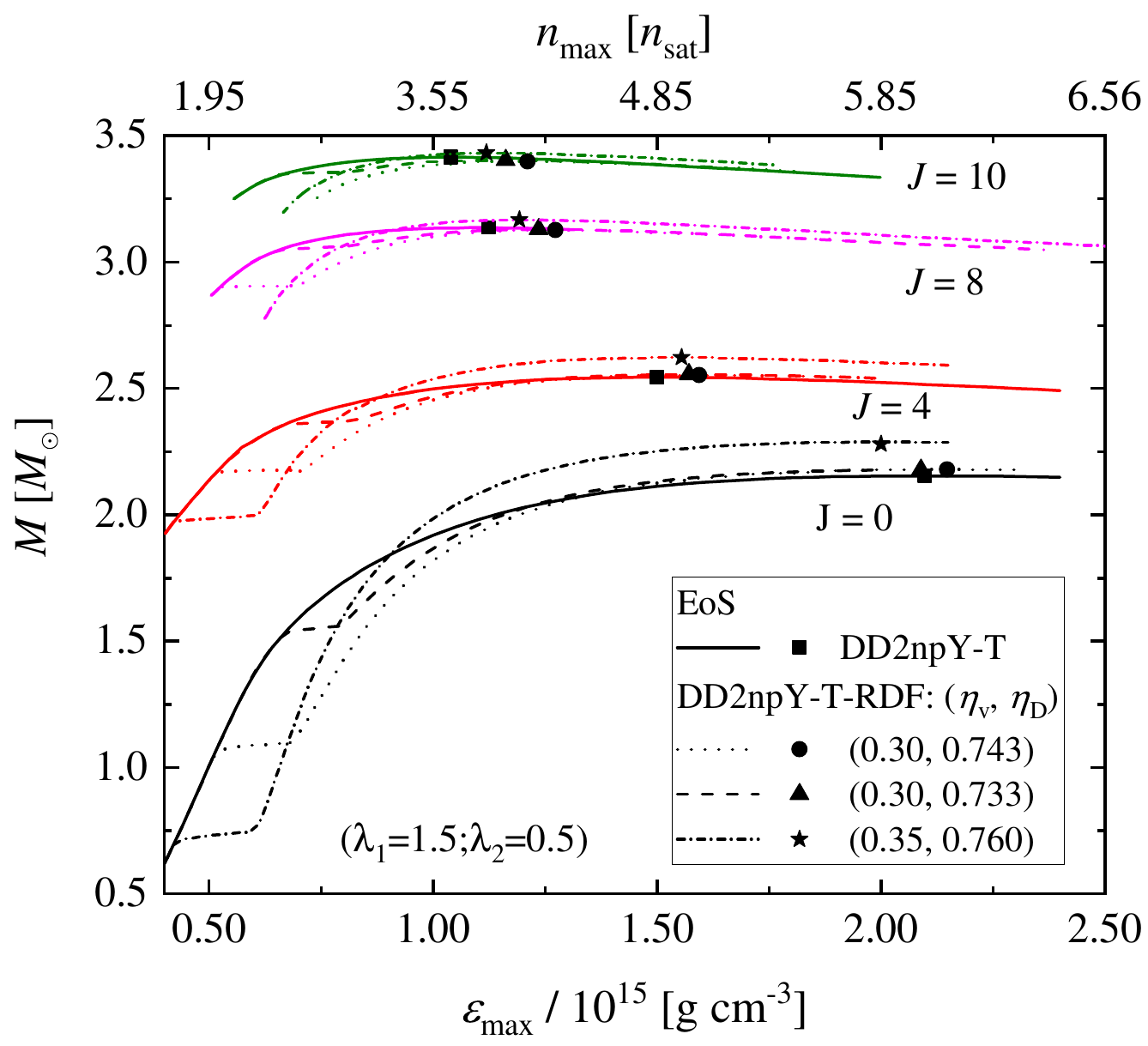}
    \includegraphics[width=0.45\textwidth]{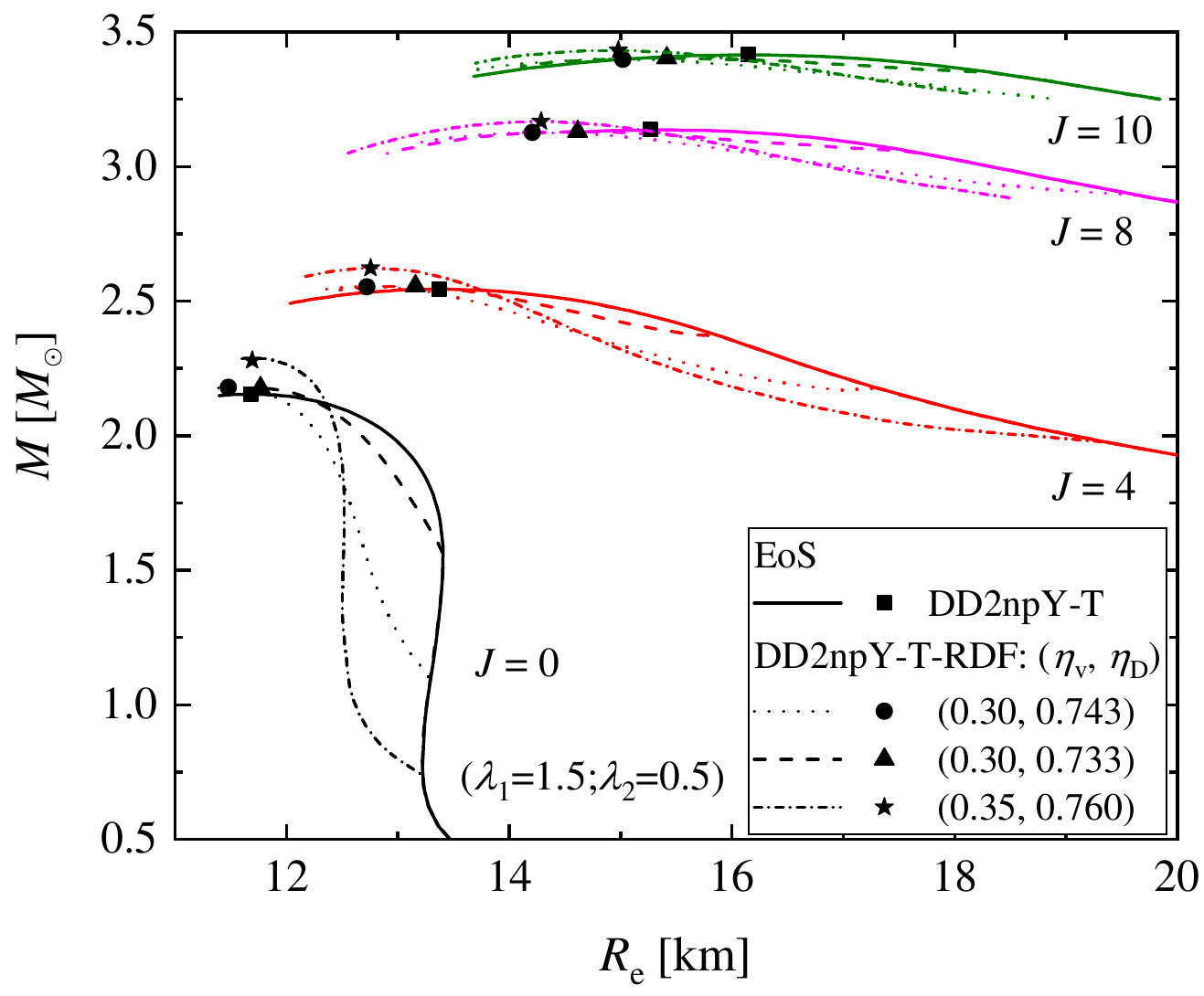}
	\caption{Gravitational mass of the quasi-toroidal configuration $(\lambda_1 = 1.5, \lambda_2 = 0.5)$ as a function of the maximum energy density. The corresponding values of the baryonic density in units of the nuclear saturation density is given on the upper $x$-axis) \textit{(upper panel)} as well as the equatorial radius of the star \textit{(bottom panel)}. The symbols indicate the maximum mass configurations.  
    }
	\label{fig:M_rho_tor}
\end{figure}
\begin{figure}[ht]
\includegraphics[width=0.45\textwidth]{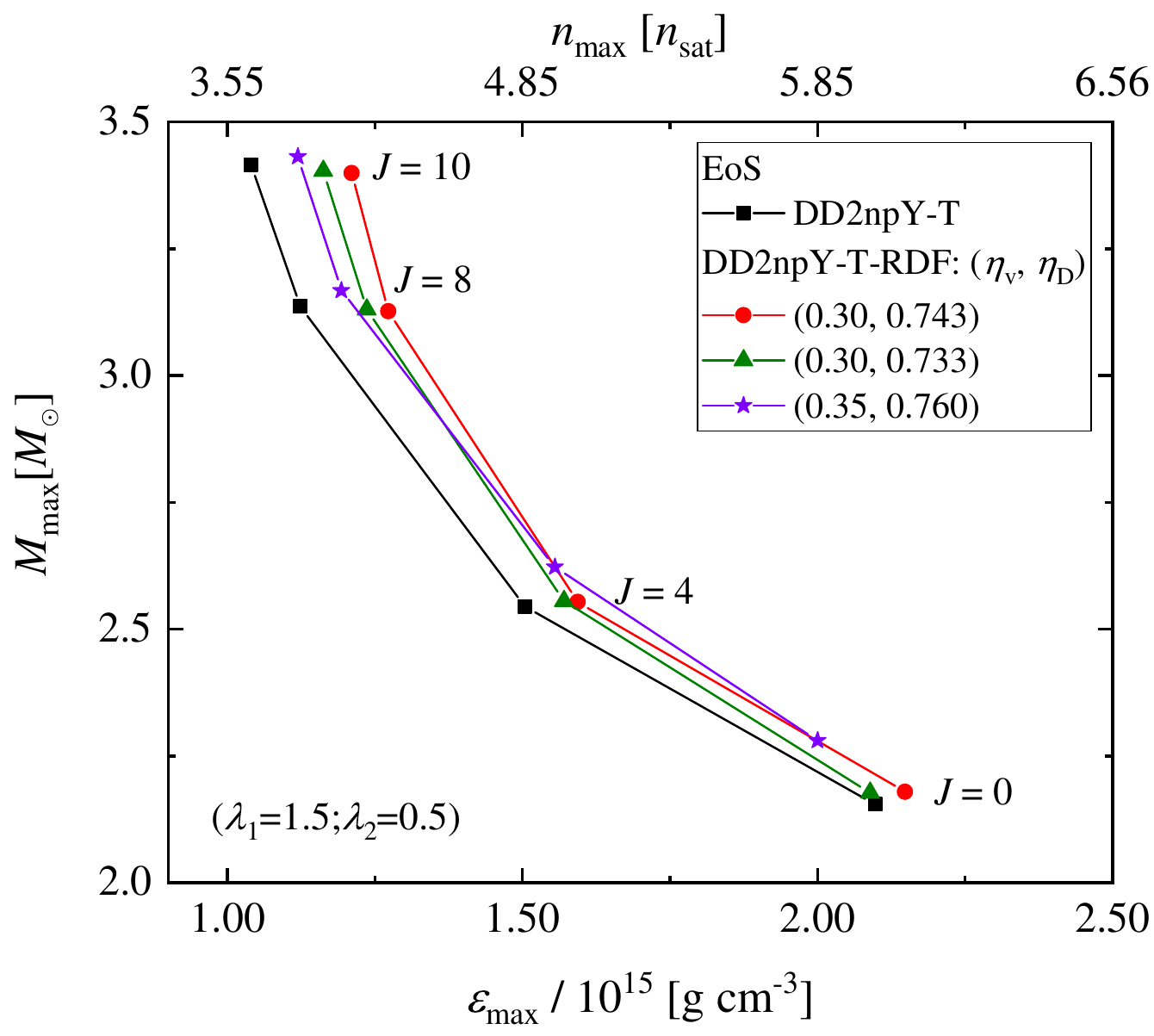}
	\caption{The maximum mass along the $J=const$ sequences as a function of the maximum energy density (lower $x$-axis) and baryonic density (upper $x$-axis) in units of the nuclear saturation density of the solutions for the sequences in Fig.~\ref{fig:M_rho_tor}.}
	\label{fig:Mmax_tor}
\end{figure}

\subsection{Quasi-toroidal (type C) solutions}
\label{sec:results_tor}

In our study, we start with the quasi-toroidal or type C solutions. For those models, we employ two pairs  $(\lambda_1 = 1.5;\lambda_2 = 0.5)$ and  $(\lambda_1 = 2.0;\lambda_2 = 0.5)$, which have already been used in the literature in GR~\cite{Iosif:2020iho, Iosif:2021-09312,Cassing:2024dxp} and modified theories of gravity~\cite{Staykov:2023ose}.  First, we concentrate on the global properties of the rotating configurations along sequences with constant angular momentum $J$ (in geometrical units $J = \frac{Jc}{GM^{2}_{\odot}}$). Since the solutions look qualitatively similar between the two pairs, in the main text we will concentrate only on  $(\lambda_1 = 1.5;\lambda_2 = 0.5)$ and some of the main results for  $(\lambda_1 = 2.0;\lambda_2 = 0.5)$ will be presented in the Appendix~\ref{app:A}. 

In the top panel of Fig.~\ref{fig:M_rho_tor} we present the gravitational mass of an NS as a function of both the maximum energy density and the baryonic density. In the bottom panel of the same figure, we present the star mass as a function of the equatorial radius for the three EoS parameterizations. The rotation rate spans between static models with $J=0$ up to $J=10$. With the increase of $J$ the sequences get shorter. On one hand this is due to the fact that lower central energy density models can typically sustain less angular moment. On the other hand, the high-$J$ solutions are numerically more challenging and the employed {\tt RNS} code is not capable to converge to them. That is why we refrain from exploring higher angular momenta in order to keep the solution sequences long enough so that the onset of the phase transition can be observed for the three studied hybrid EoSs. For the chosen maximum value of the angular momentum $J=10$, for example, the onset of the phase transition for the DD2npY-T-RDF EoS with  $(\eta_V = 0.30,\eta_D = 0.743)$ coincides with the low energy density end of the hadronic branch, below which we were not able to obtain numerical solutions due to the aforementioned limitation of the code employed. The onset for DD2npY-T-RDF EoS with $(\eta_V = 0.35,\eta_D = 0.760)$ is even lower. Hence, in those cases, the ${\tt RNS}$ code can converge only to solutions with maximum energy density corresponding to the color superconducting quark matter. It is interesting to point out that, despite the shortening of the branches with the increase of $J$, for all EoS parameterizations, the solution branches reach the turning point, the critical value of the gravitational mass along equilibrium sequences at fixed angular momentum or baryon mass which define the last stable solution. This is because the turning point occurs at lower energy density as $J$ increases.
\begin{figure*}[th!]
\includegraphics[width=0.95\textwidth]{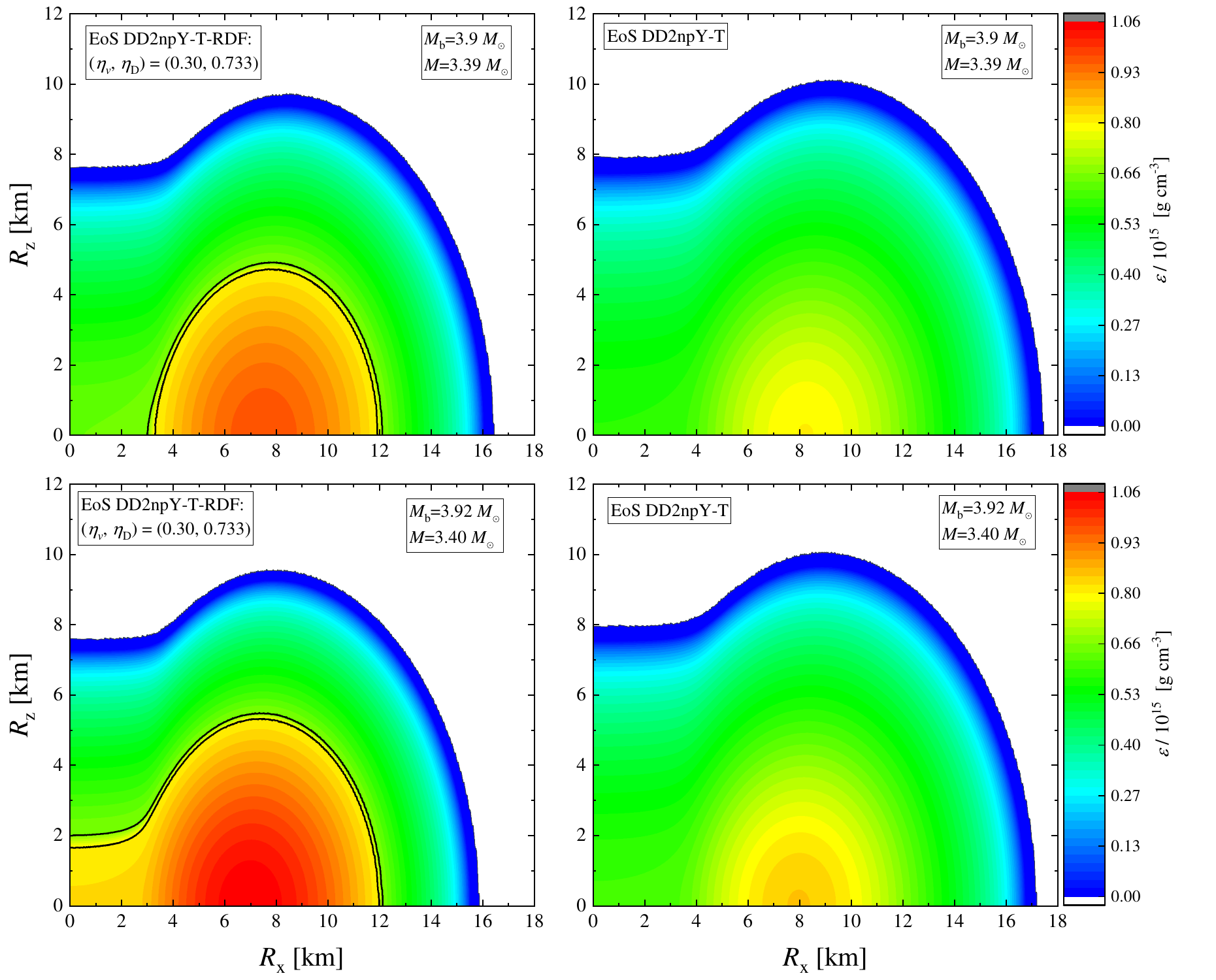}
\caption{A contour plot of the energy density distribution for the quasi-toroidal star configuration with $(\lambda_1 = 1.5, \lambda_2 = 0.5)$, fixed $J = 10$ and two different baryon masses $M_b = 3.9 \Msun$ (upper panels) and $M_b = 3.92 \Msun$ (bottom panels). The solutions in the left column are for hybrid configurations with $(\eta_V = 0.30,\eta_D = 0.733)$ and in the right column we plot pure hadronic solutions. The gravitational masses $M$ and the corresponding baryon mass $M_b$ are given in the legend. The axes represent the stellar radius in the x- and z-directions, while the color map indicates the energy density value. Hybrid star solutions are shown in the left column, whereas the right column displays purely hadronic stars. The mixed phase region is enclosed between the black lines. This is a finite region instead of an infinitely thin transition from quark to hadron matter due to a slight smoothening of the phase transition region required by the numerical code.} 
\label{fig:rho2D_tor}
\end{figure*}
Figure~\ref{fig:Mmax_tor} illustrates the observed in Fig.~\ref{fig:M_rho_tor} shift of the maximum mass solutions (the turning points along the $J=\text{const}$ sequences) toward lower maximum energy densities as $J$ increases. As it was demonstrated in~\cite{Weih:2017mcw,Lam:2025jsk,Szewczyk:2025gtz} the turning point criterion remains a sufficient condition for dynamical instability for differentially rotating stars. To the best of our knowledge, this was not studied for hybrid configurations, though we would expect the turning point criterion to be at least an approximate indicator for stability in this case as well. Although, the shift of the maximum mass is observed for both hybrid and pure hadronic EoSs, it has important implications to the hybrid case. Considering that the onset of the phase transition is at the fixed energy density, the shift of the maximum mass suggests that the span of the stable solutions with color superconducting quark matter along the $J=const$ sequence will decrease when $J$ increases. For the EoS parameterization with the late phase transitions onset and, correspondingly, higher energy density value (the DD2npY-T-RDF EoS with $(\eta_V = 0.30,\eta_D = 0.733)$), the branch shortening results in a very narrow interval of $\varepsilon_{max}$ within which potentially stable solutions exist for $J=10$. This interval is larger for the DD2npY-T-RDF EoS with $(\eta_V = 0.30,\eta_D = 0.743)$ and $(\eta_V = 0.35,\eta_D = 0.760)$, which represent the middle and early quark onset cases. It is interesting to note that, although in the static case the maximum mass corresponding to the parameter set $(\eta_V = 0.35,\eta_D = 0.760)$ is higher than for the other parameterizations considered, this difference decreases as the angular momentum $J$ increases and becomes negligible at the highest rotation rate considered, $J = 10$. According to our observations, though, this effect is not specific for the hybrid EoS and can be observed for standard hadronic ones too.
\begin{figure*}[ht]
	\includegraphics[width=0.45\textwidth]{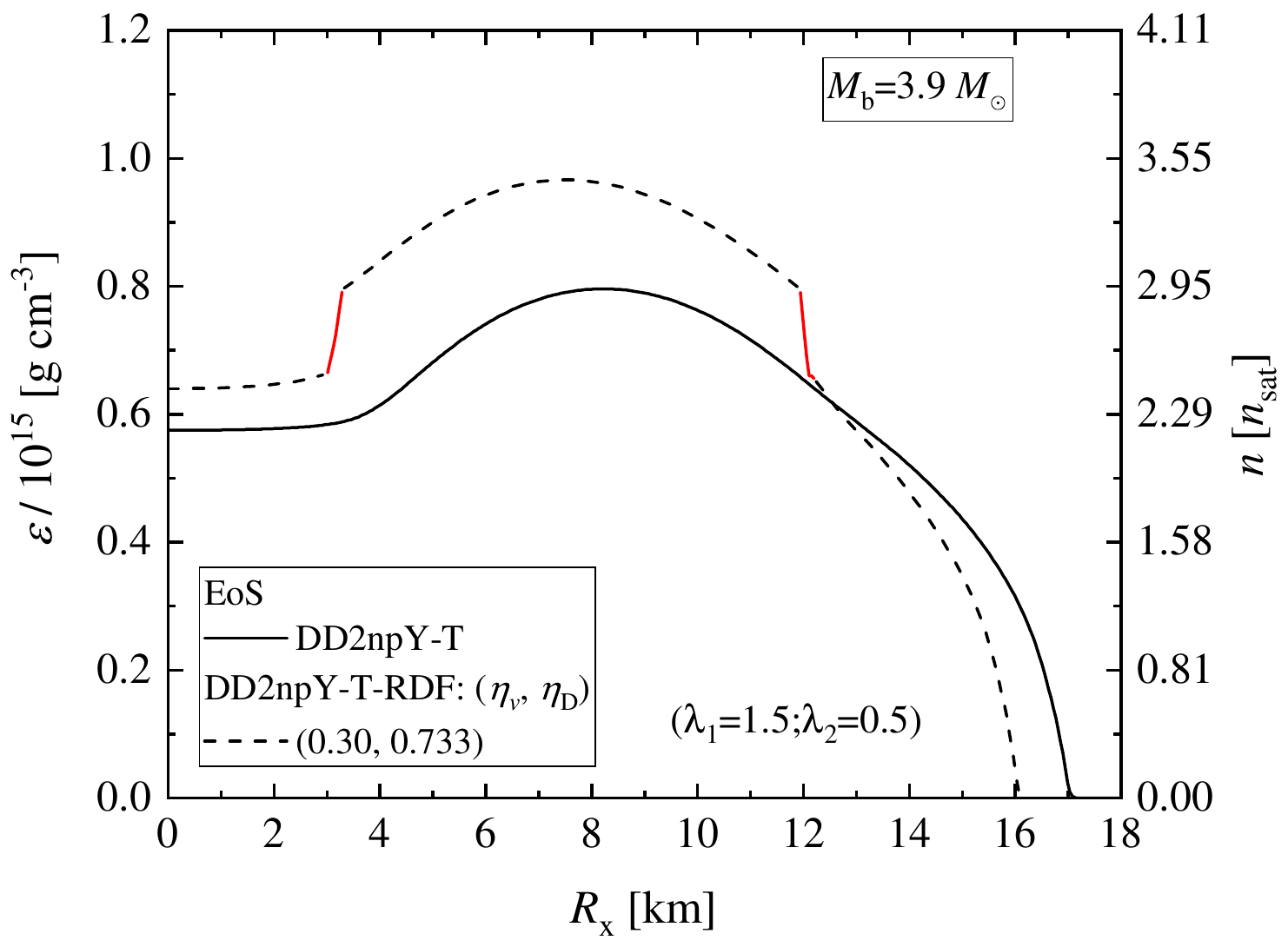}
    \includegraphics[width=0.40\textwidth]{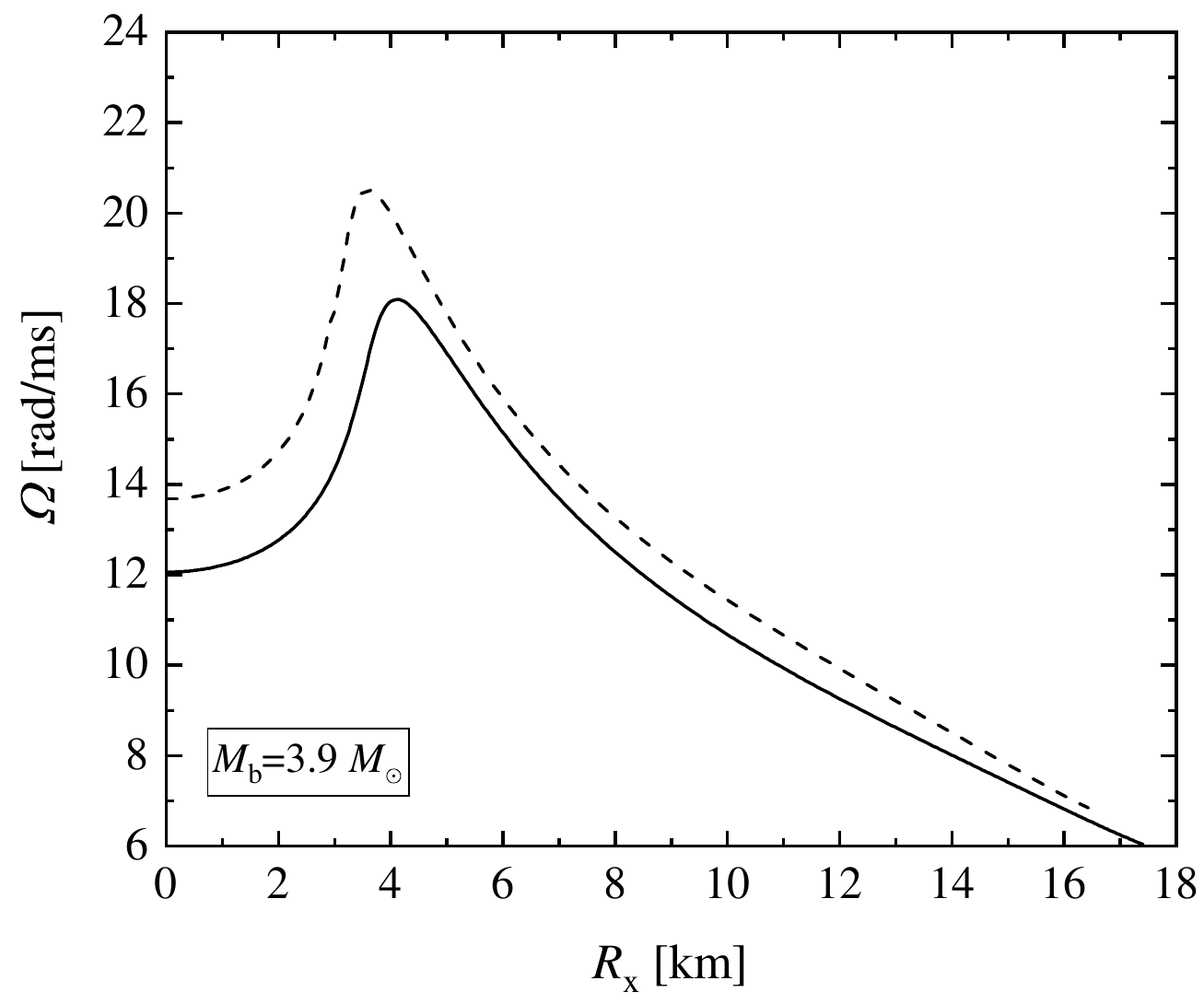}
    \includegraphics[width=0.45\textwidth]{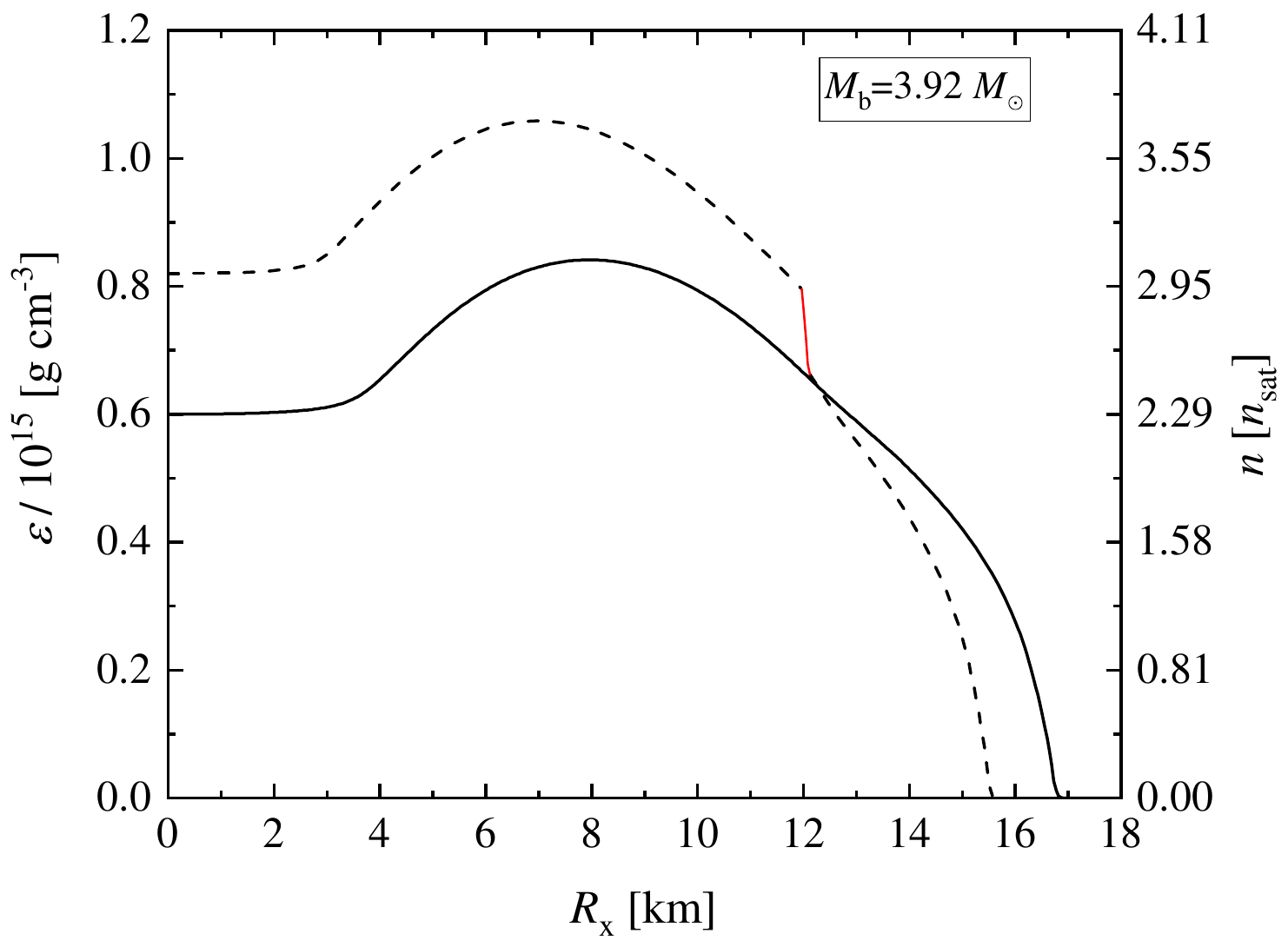}
	\includegraphics[width=0.40\textwidth]{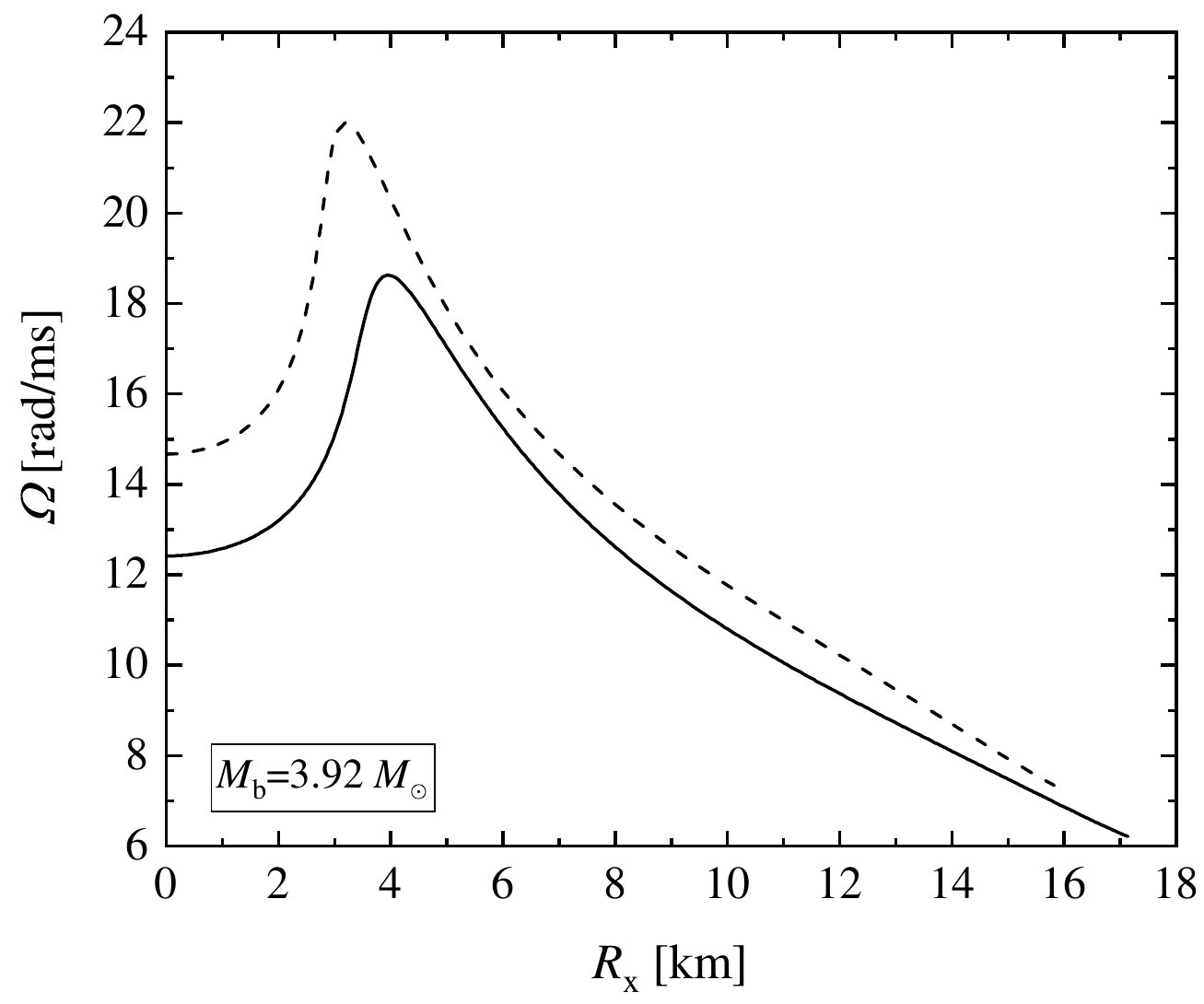}
	\caption{\textit{Left column:} The energy density profiles along the equator (the $x$-axis in Fig.~\ref{fig:rho2D_tor}) for baryon masses $M_b = 3.9 \Msun$ (upper panel) and $M_b = 3.92 \Msun$ (bottom panel) configurations in Fig.~\ref{fig:rho2D_tor}. The hadronic stars are shown with solid black curves, while the hybrid stars are depicted with black dashed curves. The mixed phase is marked with red lines. \textit{Right column:} The angular velocity profiles for the same configurations.}
	\label{fig:rho_Om_r_tor}
\end{figure*}

We proceed with a presentation of contour plots depicting the energy density profiles for purely hadronic solutions (DD2npY-T) and hybrid configurations (DD2npY-T-RDF) with $(\eta_V=0.30,\eta_D = 0.733)$. In Fig.~\ref{fig:rho2D_tor}, we compare two pairs of solutions of stars for different baryon masses. All of the solutions have fixed $J=10$. The upper and bottom panels show configurations with the baryon masses $M_b = 3.9M_\odot$ and $M_b = 3.92M_\odot$, respectively. The left column displays hybrid configurations with a maximum energy density after the quark onset, while the right column shows purely hadronic configurations before the onset. The mixed hadronic-quark phase is enclosed between two black lines and, as discussed above, it is finite due to the slight smoothening of the phase transition region required by the numerical code. In the left column in Fig.~\ref{fig:rho_Om_r_tor} we plot the equatorial radial profile of the energy density for the same models presented in Fig.~\ref{fig:rho2D_tor}. In the top panel, we show the distribution for the solution with baryon mass $M_b = 3.9 \Msun$, while in the bottom panel we plot the one for $M_b = 3.92 \Msun$. The mixed phase is marked with a red line. The radial coordinate employed on the figures is the circumferential radius defined as $R = re^{\frac{\gamma - \rho}{2}}$.

Due to the nature of the quasi-toroidal solutions, along the $J=const$ sequence, some very interesting changes in the structure of the star may occur. For sufficiently high rotation rates, we obtain purely toroidal configurations in which the stellar center and outer layers remain hadronic, with an energy density below the onset of the phase transition, while deconfined quark matter forms a toroidal structure surrounding the center (see the upper left panel of Fig.~\ref{fig:rho2D_tor}). If the central energy density of the star is above the phase transition onset, the quark matter forms a quasi-toroidal core (see the lower left panel of Fig.~\ref{fig:rho2D_tor}). In both cases, the presence of the quark phase makes the star slightly more compact compared to the pure hadronic one.

The finite area between the two black curves in the left panels of Fig.~\ref{fig:rho2D_tor}, as well as the slight inclination of the red lines in the energy density profiles (see both panels in the left column in Fig.~\ref{fig:rho_Om_r_tor}), reflect the smoothing of the EoS discussed earlier. Nevertheless, the mixed phase region remains narrow in both static and rapidly rotating models and does not affect the macroscopic properties of the star.

In the right column in Fig.~\ref{fig:rho_Om_r_tor}, we show the angular velocity profiles in the equatorial plane for the configurations displayed in Fig.~\ref{fig:rho2D_tor}. In both cases, the hybrid configurations (see the dashed curves in the right column in Fig.~\ref{fig:rho_Om_r_tor}) have a higher angular velocity compared to the purely hadronic ones (the solid curves in the right column in Fig.~\ref{fig:rho_Om_r_tor}). Despite the jump in the energy density distribution in the hybrid configuration, the angular velocity profiles remain smooth. 

\subsection{Quasi-spherical (type A) solutions}
\label{sec:results_sph}

We proceed with our study of the quasi-spherical (type A) solutions.  The recent simulations~\cite{Hanauske:2016gia,DePietri:2019mti} show that in BNS merger remnants with realistic EoS, $\lambda_1$ should be in the interval $[1.7;1.9]$ and $\lambda_2 = 1$ is favored. Just like in the quasi-toroidal case, we chose two pairs of parameters $(\lambda_1,\lambda_2) = (1.8,1.0)$ and $(\lambda_1,\lambda_2) = (2.0,1.0)$ which  we study.  In the main text of the paper we present the case $(\lambda_1,\lambda_2) = (1.8,1.0)$ and the other one is given in the appendix. In all cases, we observe numerical difficulties in finding the solutions similar to the ones described in~\cite{Iosif:2021-09312}.
\begin{figure}[ht]
     \includegraphics[width=0.45\textwidth]{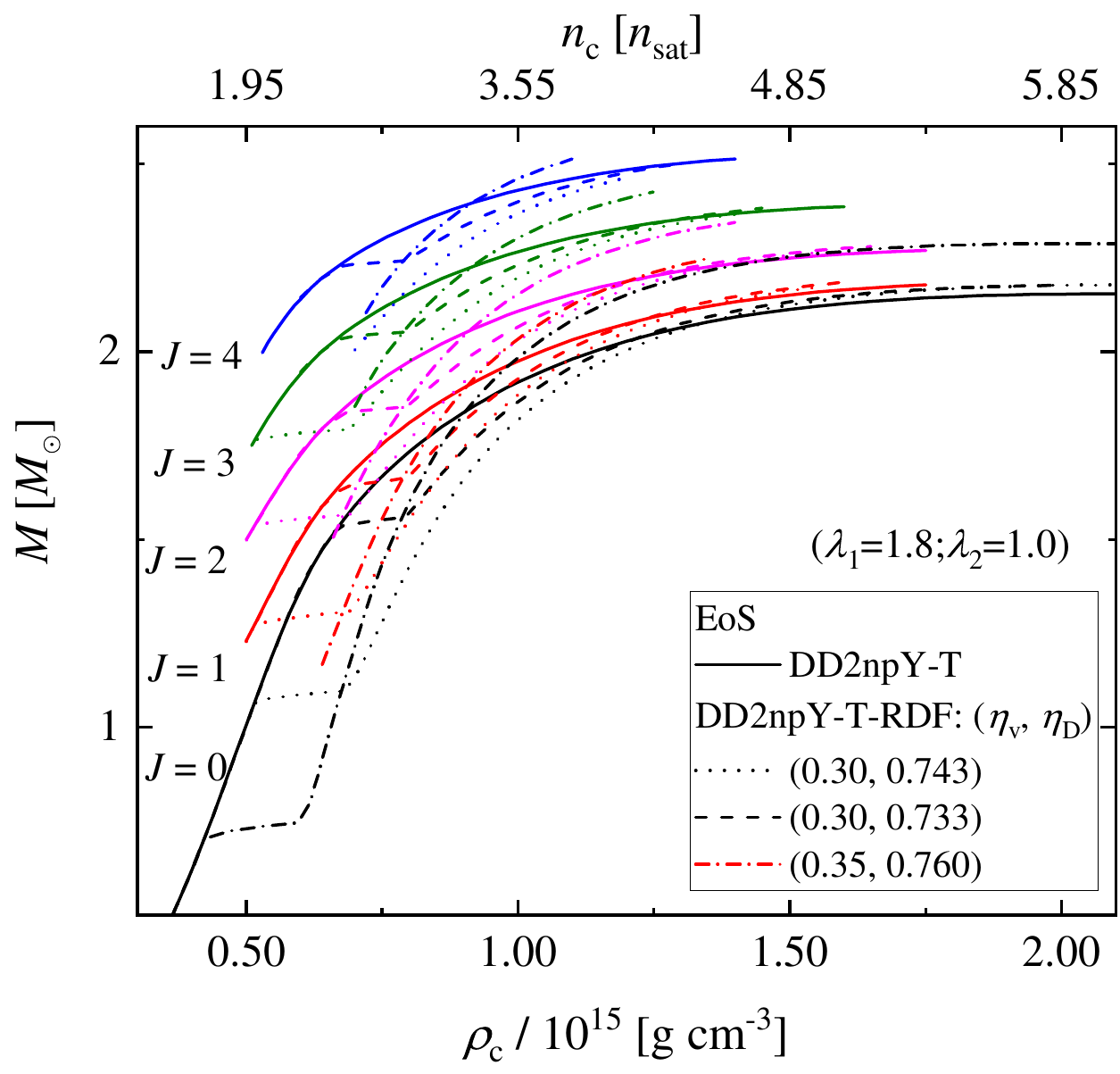}
   	\includegraphics[width=0.45\textwidth]{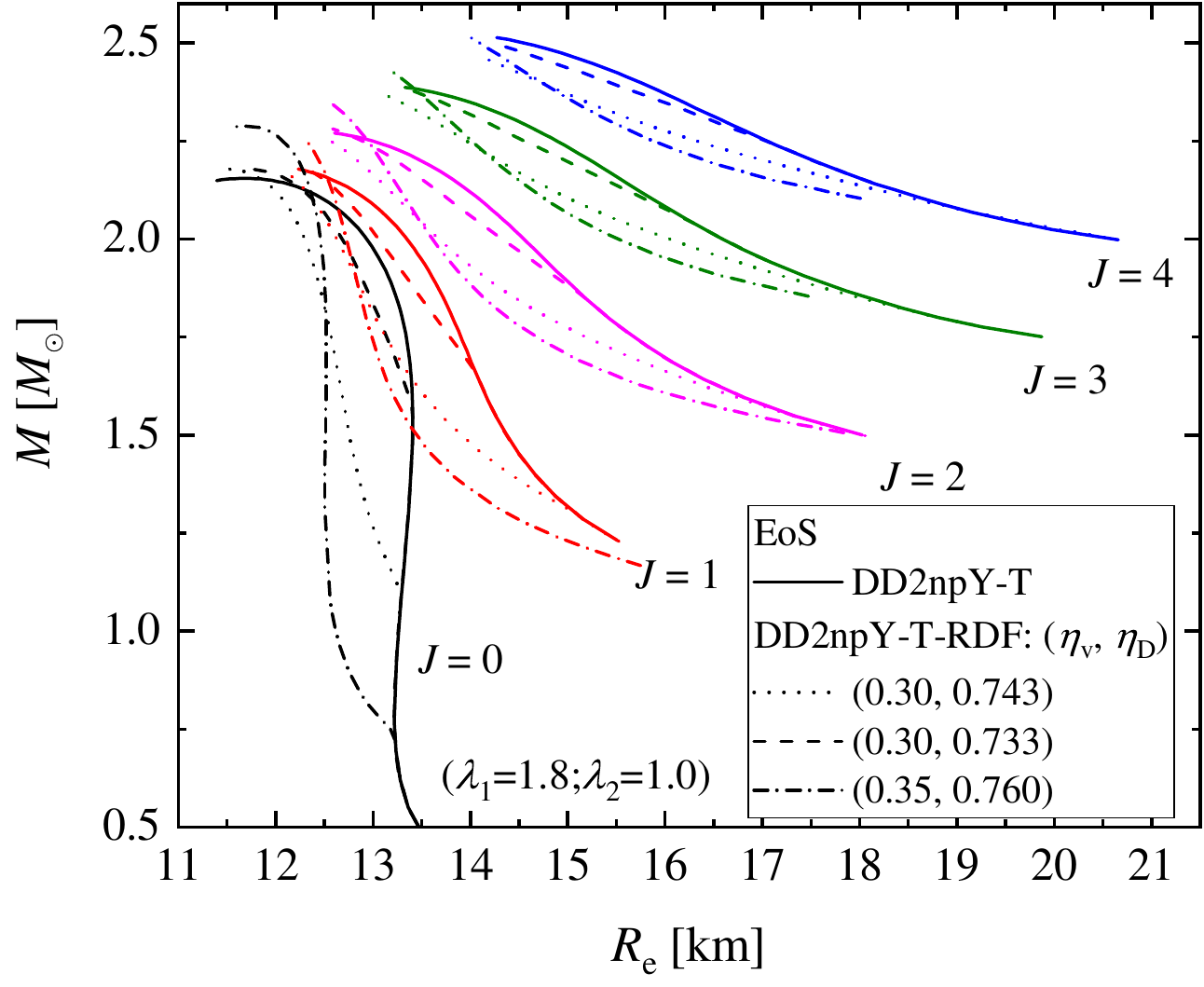}
	\caption{The gravitational mass of the quasi-spherical configurations $(\lambda_1 = 1.8, \lambda_2 = 1.0)$ as function of the central energy density  \textit{(upper panel)} and the equatorial radius \textit{(bottom panel)}. The top panel also shows the central baryonic density values on the upper x-axis. For the quasi-spherical configuration, the maximum energy density is always in the center. }
	\label{fig:M_rho_sph}
\end{figure}
\begin{figure}[ht]
    \includegraphics[width=0.48\textwidth]{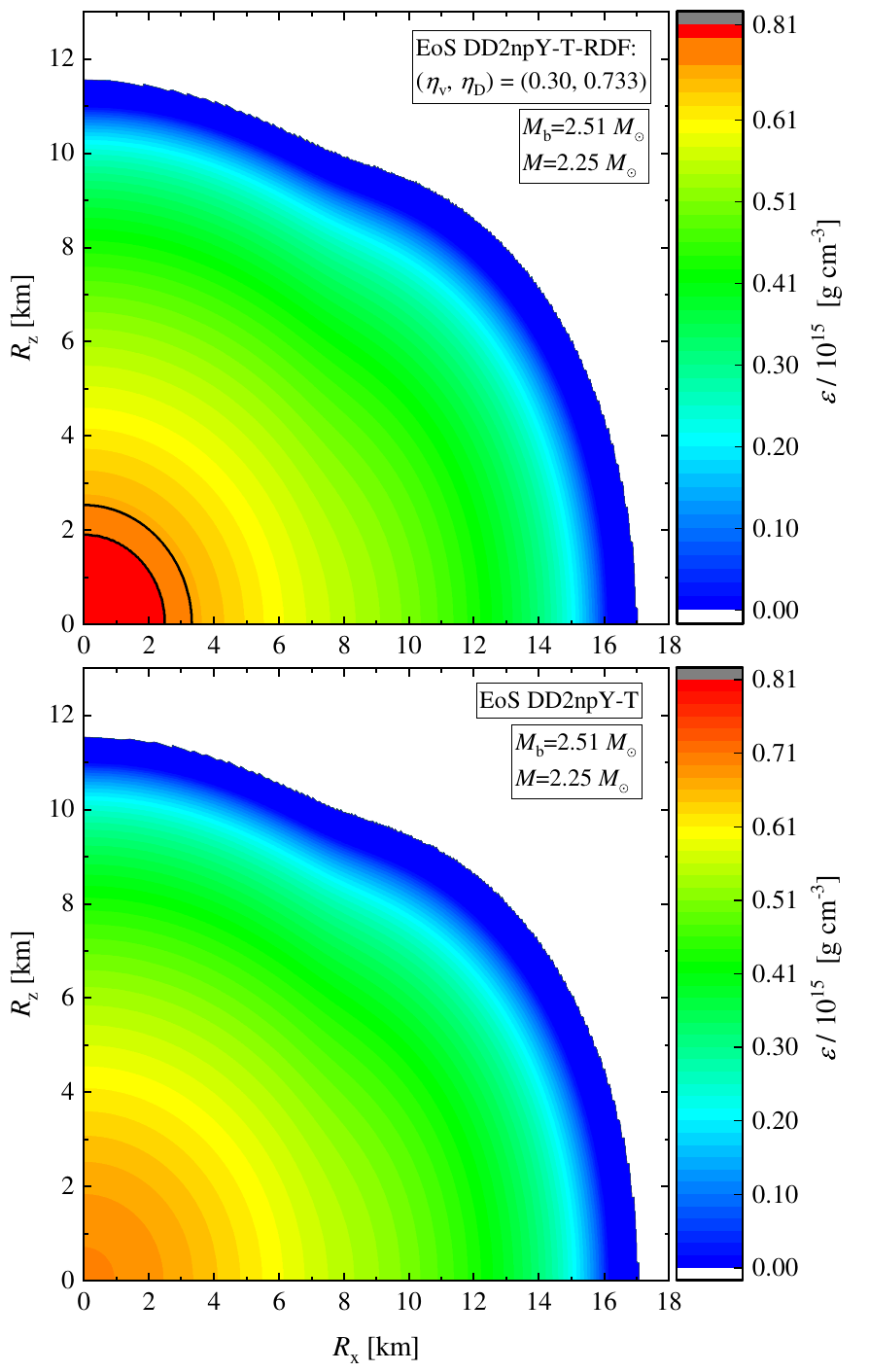}
	\caption{A contour plot of the energy distribution in stars with the quasi-spherical configuration, $(\lambda_1 = 1.8, \lambda_2 = 1.0)$, for models with $J = 4$ and the same baryon mass $M_b = 2.51 \Msun$. The corresponding gravitational masses are given on the figures as well. The solutions for the hybrid EoS and pure hadronic matter are shown on the upper and bottom panels, respectively. The mixed hadron-quark phase is enclosed between the black lines on the upper panel. }
	\label{fig:rho2D_sph}
\end{figure}
\begin{figure}[ht]
	\includegraphics[width=0.45\textwidth]{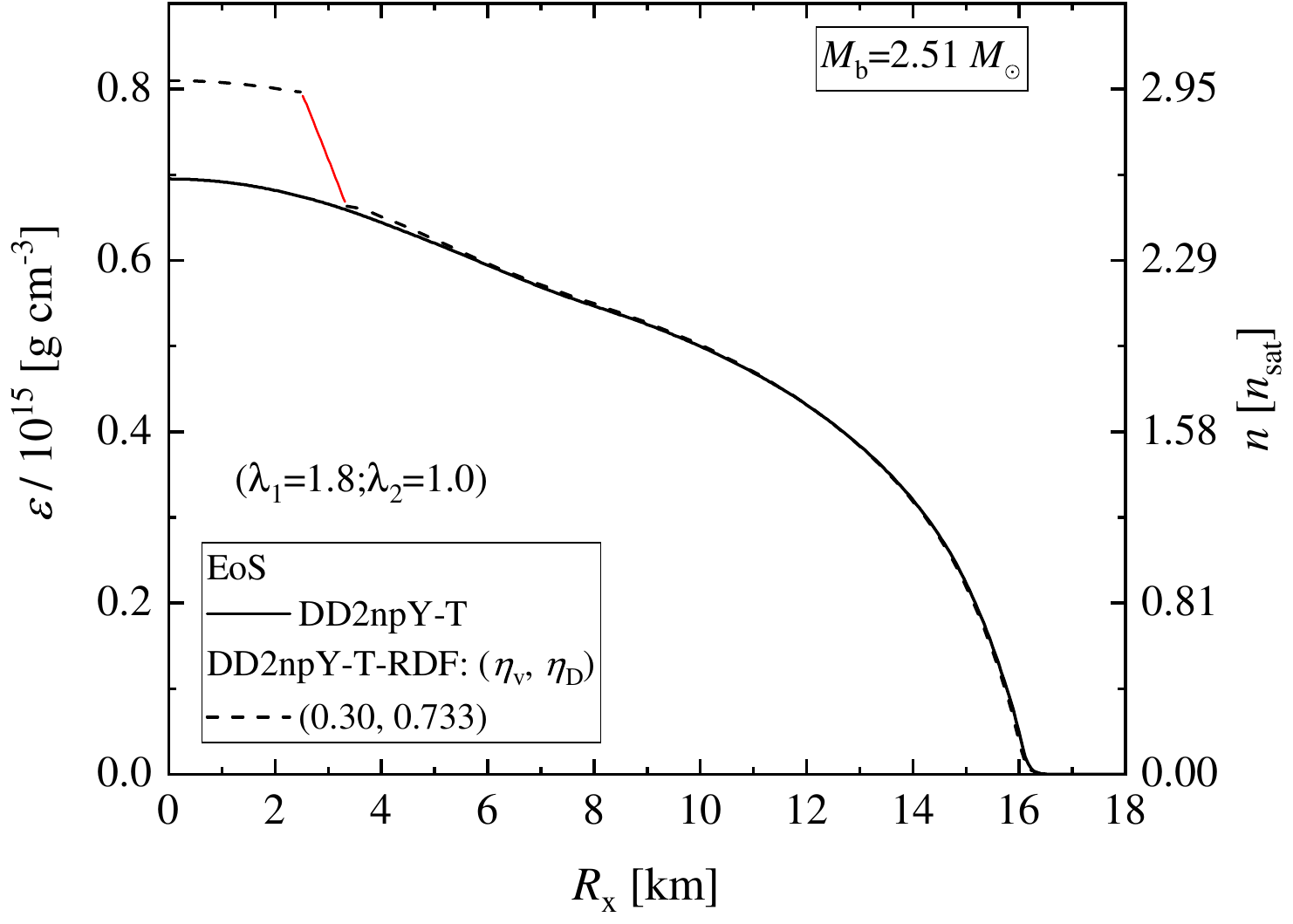}
	\caption{The same as Fig.~\ref{fig:rho_Om_r_tor}
    but for the quasi-spherical configurations.}
	\label{fig:rho_r_sph}
\end{figure}

In the top panel of Fig.~\ref{fig:M_rho_sph}, we plot the gravitational mass of the star as a function of the central energy density and the stellar radius. Note, that for the quasi-spherical stars, the maximum energy density is at the center. The set of EoSs is the same as before, and the angular momentum spans between static stars with $J=0$ and rotating solutions with $J=4$. The moderate maximum $J$, compared to the quasi-toroidal configurations, is related to numerical difficulties in obtaining quasi-spherical models~\cite{Iosif:2020iho,Iosif:2021aum}.  For example, the ${\tt RNS}$ code cannot converge for solutions with energy densities below the phase transition threshold for $J=4$ and $(\eta_V = 0.30,\eta_D = 0.743)$  and almost all of the differentially rotating sequences are terminated before reaching the $J=const$ turning point. Even though a maximum was not reached for the $J>0$ sequences in Fig.~\ref{fig:M_rho_sph} due to numerical limitations, the calculated models indicate that, similar to the type C case, the turning point and thus the termination of the stable sequence will shift to lower central energy densities as $J$ increases. 

In Fig.~\ref{fig:rho2D_sph} we compare the energy density contour plots for a pair of quasi-spherical solutions with fixed baryon mass $M_b = 2.51 \Msun$ and $J=4$. Similarly to the quasi-toroidal case, we display a solution for the hadronic (bottom panel) and hybrid configuration with $(\eta_V = 0.30,\eta_D = 0.760)$  (upper panel). For these solutions in Fig.~\ref{fig:rho_r_sph}, we plot the energy density profile in the equatorial plane. Due to the nature of the quasi-spherical solutions, presenting one pair is sufficient and the rest of the models are qualitatively similar. The reason is that for all rates of rotation, color superconducting quark matter always forms a quasi-spherical core, and no quark matter toroidal structures are possible.

\subsection{Crossing points}
\label{sec:special_points}

\begin{center}
\begin{table}[ht]
\begin{tabular}{ |c|c|c|c|c|} 
\hline
\multirow{2}{*}{CP}  &
 $M^{rot}$ & $R^{rot}$  & \multirow{2}{*}{$\eta_V$ } & \multirow{2}{*}{$\eta_D$}   \\
& [$M_{\odot}$] & [km]  & & \\
\hline
\multirow{2}{*}{A}  &
\multirow{2}{*}{2.831}  &
\multirow{2}{*}{14.791}  & 0.30 & 0.733\\ \cline{4-5}
                    & & & 0.35 & 0.760  \\ \hline
\multirow{2}{*}{B}  &
\multirow{2}{*}{2.891}                           &
\multirow{2}{*}{13.945}   & 0.35 & 0.754 \\ \cline{4-5}
                    & & & 0.35 & 0.760  \\ \hline
\multirow{2}{*}{C}  &
\multirow{2}{*}{2.853}                           &
\multirow{2}{*}{14.513}   & 0.35 & 0.760 \\ \cline{4-5}
                   & & & \multicolumn{2}{c|}{DD2npY-T} \\ \hline
\multirow{2}{*}{D}  &
\multirow{2}{*}{2.825}                           &
\multirow{2}{*}{14.725}  & 0.30& 0.733 \\ \cline{4-5}
                  & & & 0.35 & 0.760  \\ \hline
\multirow{2}{*}{E}  &
\multirow{2}{*}{2.885}                           &
\multirow{2}{*}{13.947}    & 0.35& 0.754 \\ \cline{4-5}
                  & & & 0.35 & 0.760  \\ \hline
\multirow{2}{*}{F}  &
\multirow{2}{*}{2.841}                           &
\multirow{2}{*}{14.562}    & 0.35& 0.760 \\ \cline{4-5}
                       & & & \multicolumn{2}{c|}{DD2npY-T}  \\ \hline
\multirow{2}{*}{G}  &
\multirow{2}{*}{2.371}                           &
\multirow{2}{*}{13.481} & 0.30& 0.733 \\ \cline{4-5}
                      & & &0.35 & 0.760  \\ \hline
\multirow{2}{*}{H}  &
\multirow{2}{*}{2.385}                           &
\multirow{2}{*}{13.414}      & 0.35& 0.760 \\ \cline{4-5}
                      & & &\multicolumn{2}{c|}{DD2npY-T}  \\ \hline
\end{tabular}
\caption{Parameters of the CPs. The columns represent their gravitational mass and radius, as well as the vector $\eta_V$ and diquark $\eta_D$ couplings of the corresponding mass-radius curves.}
\label{tbl:SPs}
\end{table}
\end{center}
\endgroup
\begin{figure}[ht]
	\includegraphics[width=0.44\textwidth]{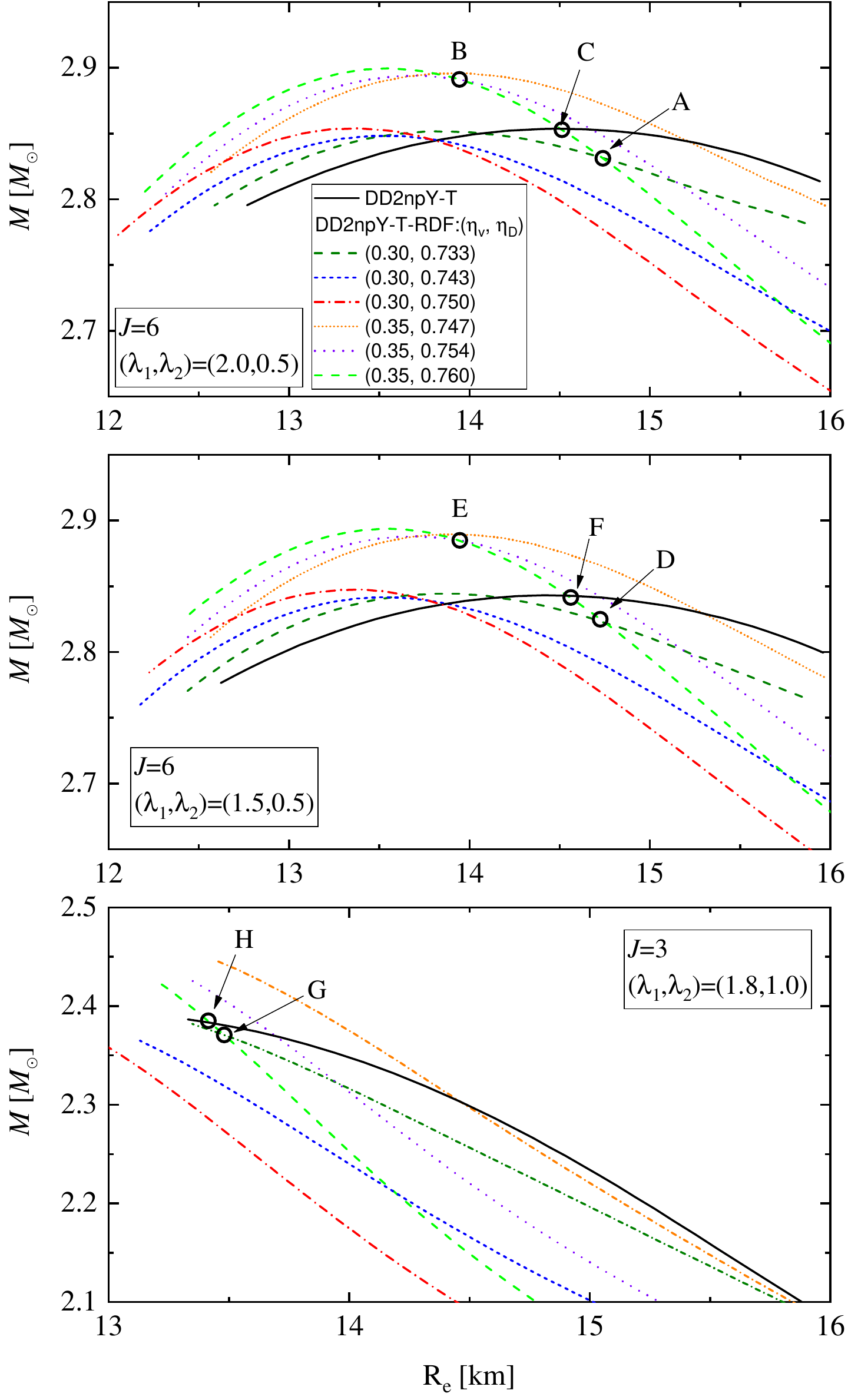}
	\caption{Mass--radius curves around the CPs. The examined models are marked with circles. Only one pair $(\lambda_1,\lambda_2)$ is presented in the quasi-spherical case because for the second considered pair the sequences are too short.
	}
	\label{fig:special_MR}
\end{figure}
\begin{figure*}[ht]
	\includegraphics[width=0.45\textwidth]{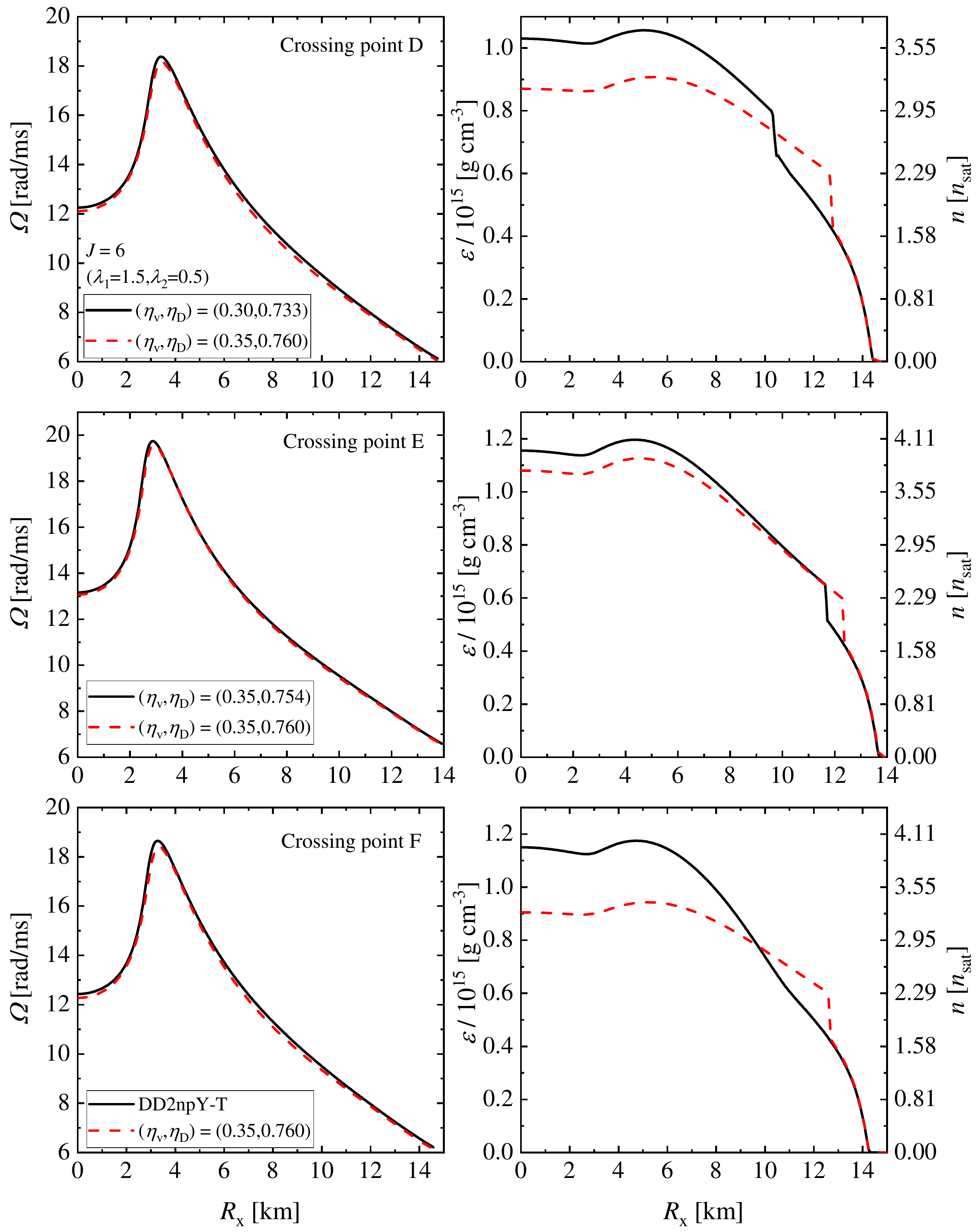}
	\includegraphics[width=0.45\textwidth]{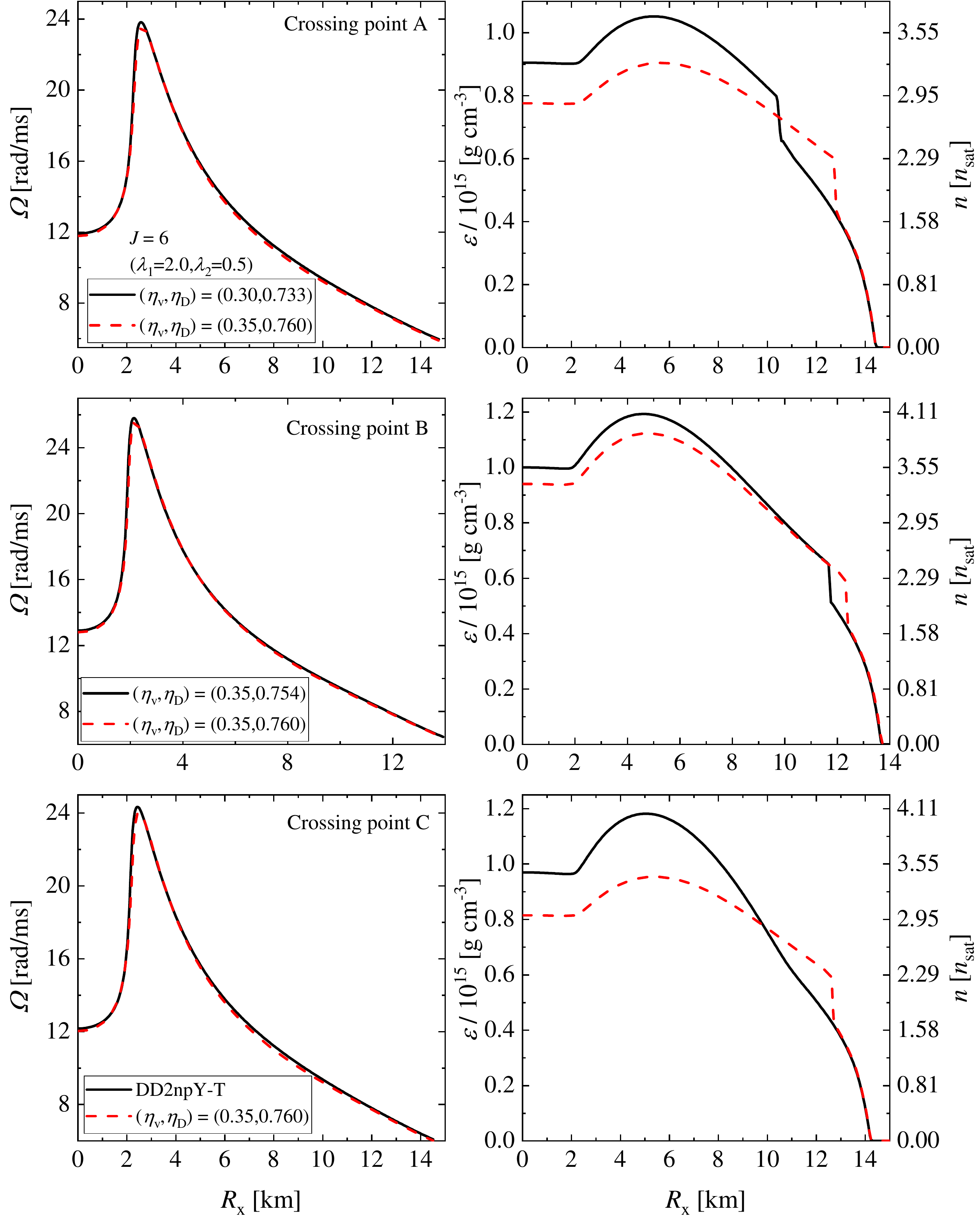}
	\caption{The angular velocity and the energy density profiles of stars in the equatorial plane for the quasi-toroidal solutions with $J=6$ in three different CPs. \textit{Left} $(\lambda_1,\lambda_2) = (1.5,0.5)$ and \textit{right} $(\lambda_1,\lambda_2) = (2.0,0.5)$. The capital letters on the panels correspond to the CPs presented in Fig.~\ref{fig:special_MR} and Table~\ref{tbl:SPs}. 
	}
	\label{fig:special_tor}
\end{figure*}
\begin{figure*}[ht]
	\includegraphics[width=0.6\textwidth]{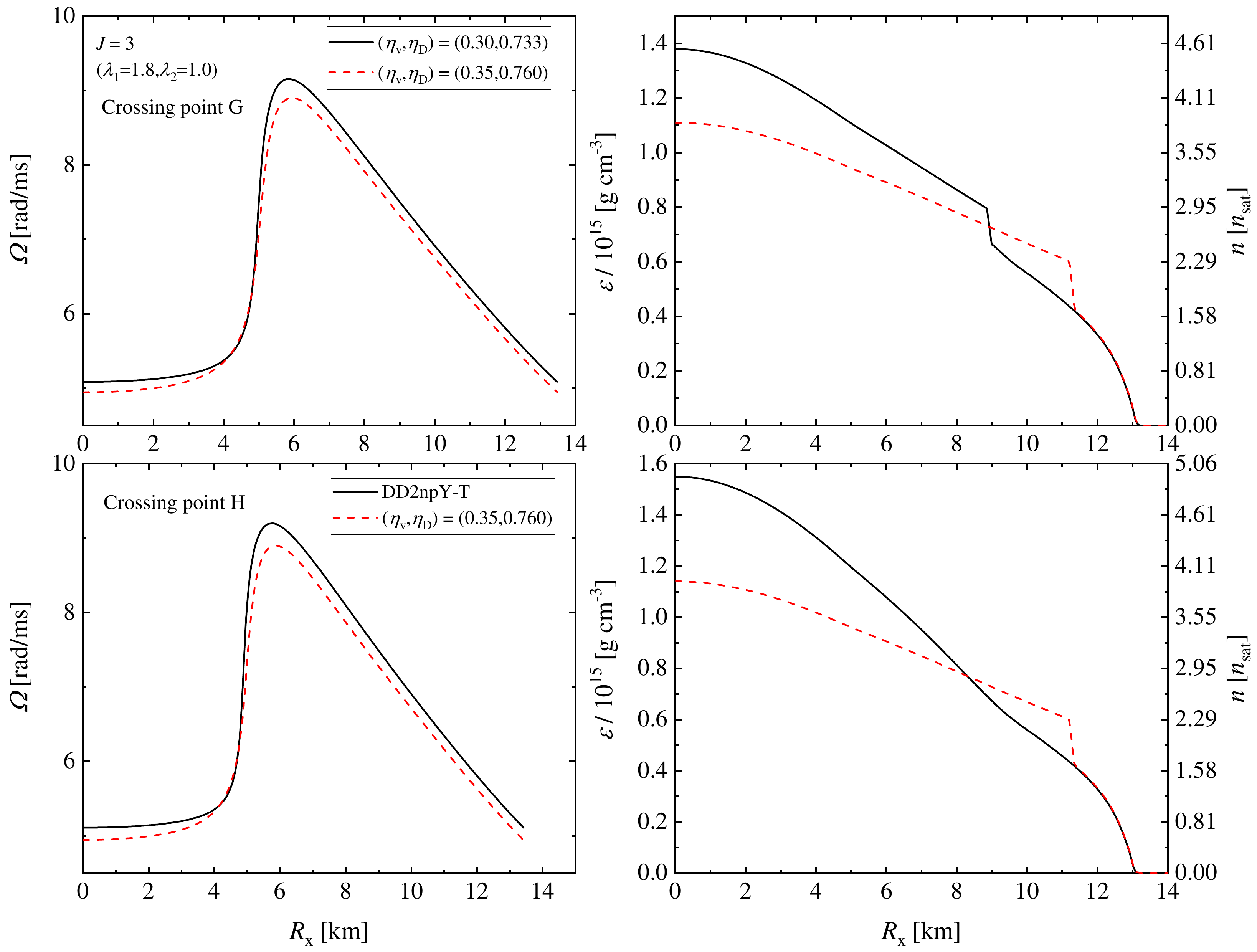}
	\caption{The angular velocity and the energy density profiles in the equatorial plane for the quasi-spherical solutions with $J=3$. Only two CPs for the pair $(\lambda_1,\lambda_2) = (1.8,1.0)$ are considered due to the disappearance of the CPs following the shortening of the branches. The capital letters on the panels correspond to the CPs presented in Fig.~\ref{fig:special_MR} and Table~\ref{tbl:SPs}.
	}
	\label{fig:special_sph}
\end{figure*}
To examine the internal structure of hybrid stars in greater detail, we compare differentially rotating stellar configurations located at the CPs, i.e., the intersection points of several mass–radius curves corresponding to stars with different internal compositions. In the case of hybrid stars, the configurations considered have identical masses and radii but differ in the size of their quark matter cores, the stiffness of quark matter (parametrized by $\eta_V$), and the onset density of deconfinement (parametrized by $\eta_D$). These configurations are also compared to purely hadronic ones. To study more realistic models reproducing merger remnants, we chose angular momentum $J=6$ for the quasi-toroidal case and $J=3$ for the quasi-spherical case. The two values are motivated by BNS merger simulation results~\cite{Uryu:2017obi, Xie:2020udh}. 

We study the CPs presented in Table~\ref{tbl:SPs} for three of the four pairs of parameter $(\lambda_1, \lambda_2)$ mentioned in the previous sections. For the quasi-toroidal case, we consider $J=6$ and both $(\lambda_1,\lambda_2) = (1.5,0.5)$ and $(\lambda_1,\lambda_2) = (2.0,0.5)$. The resulting mass-radius curves are presented in Fig.~\ref{fig:special_MR}, while the CPs are marked by circles and labeled. The angular velocity and energy density profiles for the selected configurations are shown in Fig.~\ref{fig:special_tor}. The left panel corresponds to the differential rotation law with $(\lambda_1,\lambda_2) = (1.5,0.5)$, while the right panel shows the results for $(\lambda_1,\lambda_2) = (2.0,0.5)$. The same set of configurations is considered in both panels. In the first two rows of Fig.~\ref{fig:special_tor} we compare only hybrid configurations. In the first case both $\eta_V$ and $\eta_D$ are different between the compared models, while in the second case $\eta_V$ is the same and only $\eta_D$ is different. In the third row we compare a hybrid configuration with pure hadronic NS. Interestingly, despite the obvious differences in the energy density profiles, the angular velocity distributions are almost identical for all studied pairs of models. 

For the quasi-spherical solutions with $J=3$ some of the branches are terminated before they intersect, and hence the CP cannot be obtained. This is the case for both pairs $(\lambda_1,\lambda_2)$ but it is more extreme for the set $(\lambda_1,\lambda_2) = (2.0,1.0)$. Hence, in Fig.~\ref{fig:special_sph} we present only two CPs for $(\lambda_1,\lambda_2) = (1.8,1.0)$. The mass--radius curves are in the middle panel of Fig.~\ref{fig:special_MR} and the CPs are marked with circles and labeled.  

The discrepancies between the energy density distributions in Fig.~\ref{fig:special_sph} are more significant, compared to the quasi-toroidal case, and we can see minor deviation between the angular velocity profiles. Still, the latter remain very close for models with the same baryon mass and angular momentum. 

We want to emphasize that the observed degeneracy between the rotational profiles is based on a single differential rotation law with a specific set of EoSs. Therefore, more detailed exploration of the parameter space is needed to determine up to what extent this degeneracy holds.

\section{Conclusions}
\label{sec:conclusions}

In the present paper, we studied the global characteristics and the internal structure of differentially rotating hybrid stars with a deconfinement phase transition from hadronic to color superconducting quark matter. The hadronic phase is described by the DD2npY-T EoS, which includes not only nucleonic degrees of freedom but also hyperons. Quark matter is modeled using the RDF approach with medium-dependent scalar and pseudoscalar couplings, and includes the 2SC color-superconducting phase. The quark matter model is characterized by two key parameters: the dimensionless vector coupling $\eta_V$ and the diquark pairing strength $\eta_D$. For the vector coupling $\eta_V$ we set two different values and six different values for the diquark pairing strength $\eta_D$. These parameters are combined into six distinct pairs, yielding six EoS parameterizations that systematically explore different hybrid star properties. The exact values and combinations of those parameters are given in Table~\ref{tbl:Param_table}.  In particular, we have configurations with early, intermediate and late deconfinement onsets and two different stiffnesses at each onset. 

The differential rotation is modeled with the realistic four-parameter law by Ury\={u}~\cite{Uryu:2017obi} which allows the maximum of the angular velocity profile to be shifted away from the center, a characteristic observed in the BNS merger remnants obtained in full relativistic simulations. 

Within this framework, using the {\tt RNS} code and six parametrizations for the hybrid EoS, we obtained both type A (quasi-spherical) and type C (quasi-toroidal) differentially rotating solutions. The strong softening of matter at the first-order deconfinement phase transition leads to a reduction in the stellar radius; however, the stiffening of quark matter due to color superconductivity allows the maximum stellar mass to reach, and even exceed, that of purely baryonic stars.

With the increase of the maximum energy density, tending to the maximum mass solutions, the mass of hybrid stars gets similar to the mass of pure hadronic ones for $\eta_V = 0.30$ and even increase significantly for $\eta_V = 0.35$. An interesting observation for the type C solutions is the shift of the $J=const$ sequences' turning point to lower maximum energy density as $J$ increases. This implies that, if the turning point stability criterion remains valid for differentially rotating hybrid configurations, the interval of potentially stable hybrid solutions, defined as those between the phase transition onset and the turning point, shrinks with increasing $J$. However, due to limitations of the numerical code, most type A solutions terminate before reaching the turning point, and therefore, no definitive conclusions can be drawn in that case.

An interesting qualitative difference emerges in the internal structure of the two types of solutions. For hybrid type C solutions having maximum energy density away from the center, it is possible for the central energy density to be below the deconfinement onset. As the energy density increases in the radial direction, it can go above the onset, reaching the maximum at some point in the star, past which it decreases. For those models, we observe the formation inside the star of a toroidal structure consisting of color superconducting quark matter. For the type C models with central energy density above the onset and all hybrid type A models, a typical quark matter core is formed.  Since the employed Uryu differential rotation law can mimic well the post-merger remnant profiles, the construction of such equilibrium models is a first step towards the more important problem, namely studying their oscillation spectrum which will reveal itself in post-merger GW signal. Only such approximate methods can allows us to make a extensive parameteric study of some post-merger properties. Interestingly, despite the presence of abrupt changes in the stellar density profile, the rotational profiles remain smooth with no clear signatures related to the phase transition.

The set of EoS parameterizations we study allows for the intersections of the mass-radius curves for different parameter sets. Those CPs are similar to the special points in the static case. We studied two cases -- quasi-toroidal models with $J=6$ and quasi-spherical models with $J=3$. Comparing the solutions at the CPs, it turns out that the angular velocity profiles are identical despite the somewhat large difference in the energy density distributions and the underlying EoS. This result demonstrates that similar post-merger remnant properties can arise from models with and without phase transitions, underscoring the crucial role of multi-messenger observations in breaking this degeneracy. It should be pointed out, though, that those results are based on the single differential rotation law we adopted in this paper with a limited set of EoS and also two combinations for differential rotation parameters. Therefore, even though we have seen strong indications for a EoS independence of the rotational profiles, a more general conclusion would require performing a thorough parameter exploration involving also other rotation laws. The study represents an important step toward understanding the impact of the deconfinement phase transition on the post-merger evolution and the remnant’s ability to resist gravitational collapse.

\acknowledgments
V.S. gratefully acknowledges support from the UK Research and Innovation (UKRI)-funded ``The next-generation gravitational-wave observatory network'' project (Grant No. ST/Y004248/1). V.S. acknowledges the gratitude to the Funda\c c\~ao para a Ci\^encia e Tecnologia (FCT) I.P. for support under Advanced Computing Projects 2024.07037.CPCA.A1 with DOI identifier 10.54499/2024.07037.CPCA.A1 and 2023.10526.CPCA.A2 with DOI identifier 10.54499/2023.10526.CPCA.A2. K. V. S. and S. S. Y. acknowledge the partial support of Grant No. KP-06-N62/6 from the Bulgarian science fund. D. D. D. acknowledges financial support via an Emmy Noether Research Group funded by the German Research Foundation (DFG) under grant no. DO 1771/1-1, by the Spanish Ministry of Science and Innovation via the Ram\'on y Cajal programme (grant RYC2023-042559-I), funded by the Spanish Ministry of Science and Innovation and the Spanish State Research Agency (MCIN/AEI/10.13039/501100011033), and by the European Social Fund Plus (ESF+), and by the Spanish Agencia Estatal de Investigaci\'on (grant PID2024-159689NB-C21) funded by the Ministerio de Ciencia, Innovaci\'on y Universidades. We acknowledge Discoverer PetaSC and EuroHPC JU for awarding this project access to Discoverer supercomputer resources. Calculations were also carried out on HEMUS supercomputer at the Center of Excellence in Informatics and ICT.

\appendix 
\label{app:A}
\section{Supplementary results for quasi-toroidal and quasi-spherical solutions}

In the discussion above, we referred to two sets of differential-rotation parameters, $(\lambda_1,\lambda_2)$, for both the quasi-toroidal and quasi-spherical configurations; however, only a subset of the results was presented and discussed. Below we present some additional results for the quasi-toroidal solutions with $(\lambda_1,\lambda_2) = (2.0,0.5)$, and for quasi-spherical solutions with $(\lambda_1,\lambda_2) = (2.0,1.0)$. 

\begin{figure}[ht]
 	\includegraphics[width=0.45\textwidth]{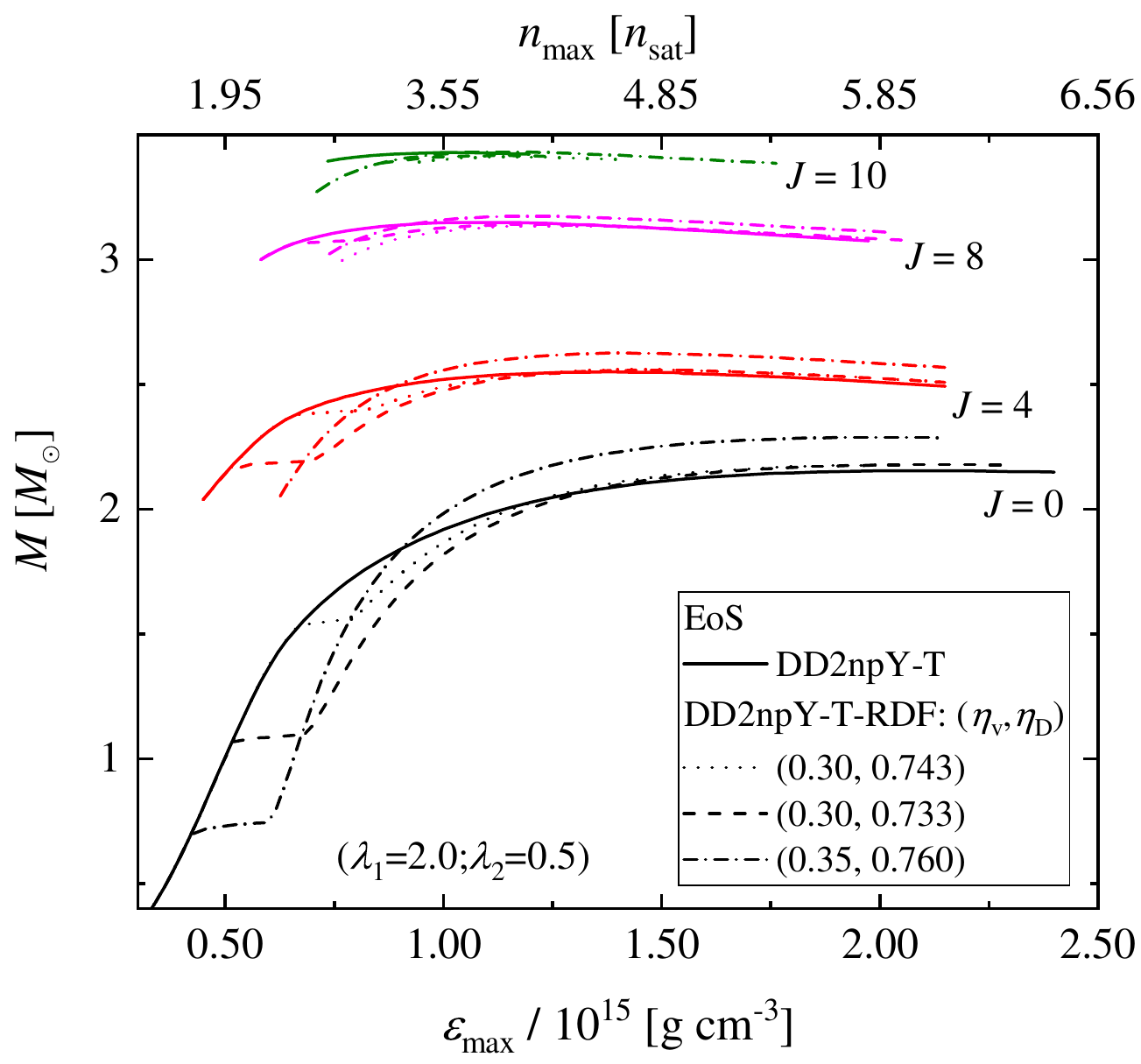}
   	\includegraphics[width=0.45\textwidth]{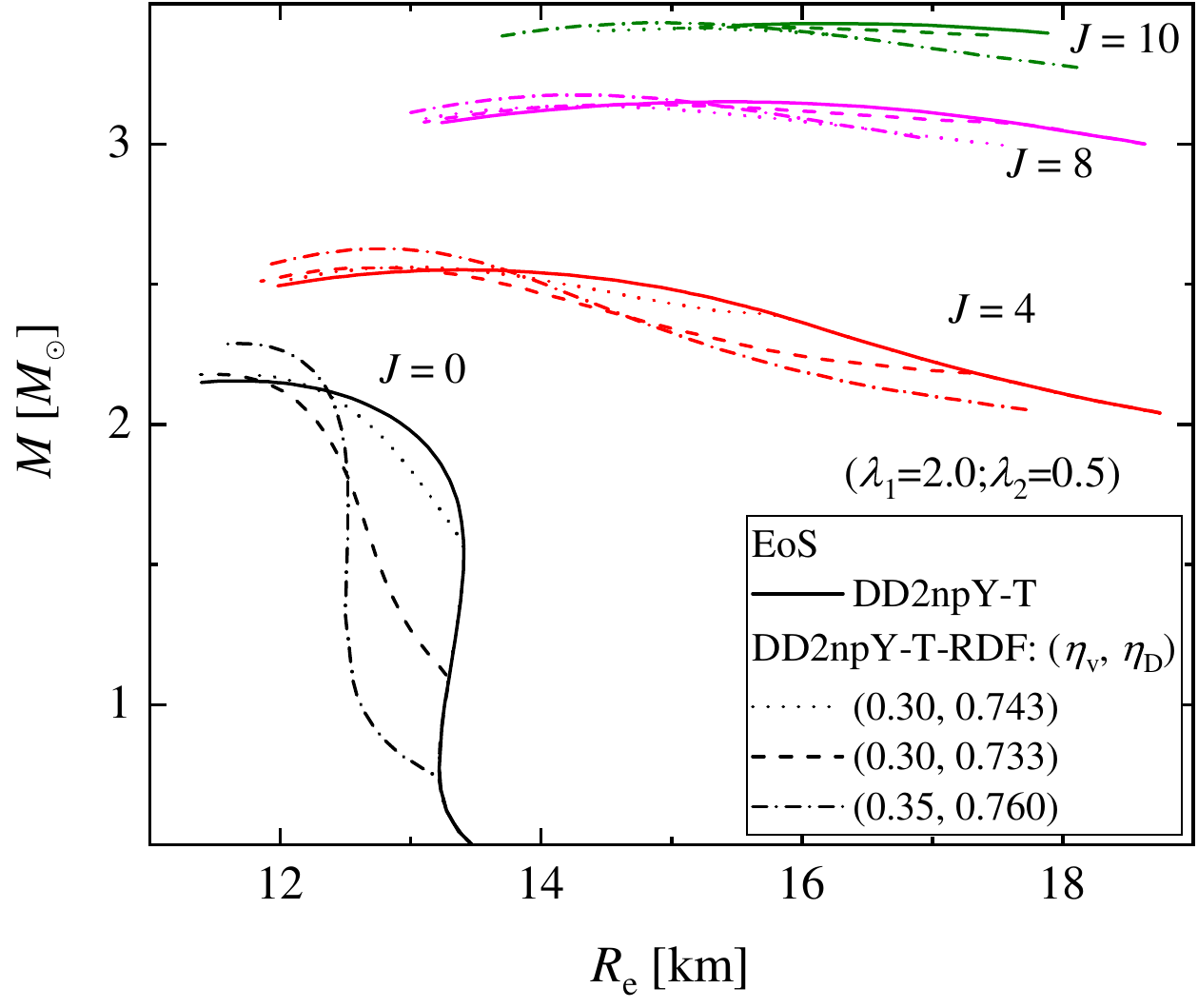}
	\caption{Quasi-toroidal solutions with $(\lambda_1, \lambda_2) = (2.0,0.5)$. All of the notations and the EoS are the same as in Fig.~\ref{fig:M_rho_tor}.
    }
    \label{fig:M_rho_tor_2}
\end{figure}

Fig.~\ref{fig:M_rho_tor_2} focuses on the quasi-toroidal solutions with $(\lambda_1,\lambda_2) = (2.0,0.5)$. The top panel depicts the mass as a function of the maximum energy density (baryon density on the top x-axis) and the bottom panel -- the mass as a function of the equatorial radius. The presented rates of rotation are up to $J=10$ just like in Fig.~\ref{fig:M_rho_tor}. The only observed difference from the case $(\lambda_1,\lambda_2) = (1.5,0.5)$ presented in Fig.~\ref{fig:M_rho_tor} is that the branches with the highest rate of rotation are slightly shorter at the low density end for $(\lambda_1,\lambda_2) = (2.0,0.5)$. Otherwise the observed behavior is qualitatively the same for both pairs $(\lambda_1,\lambda_2)$.  

\begin{figure}[ht]
\includegraphics[width=0.45\textwidth]{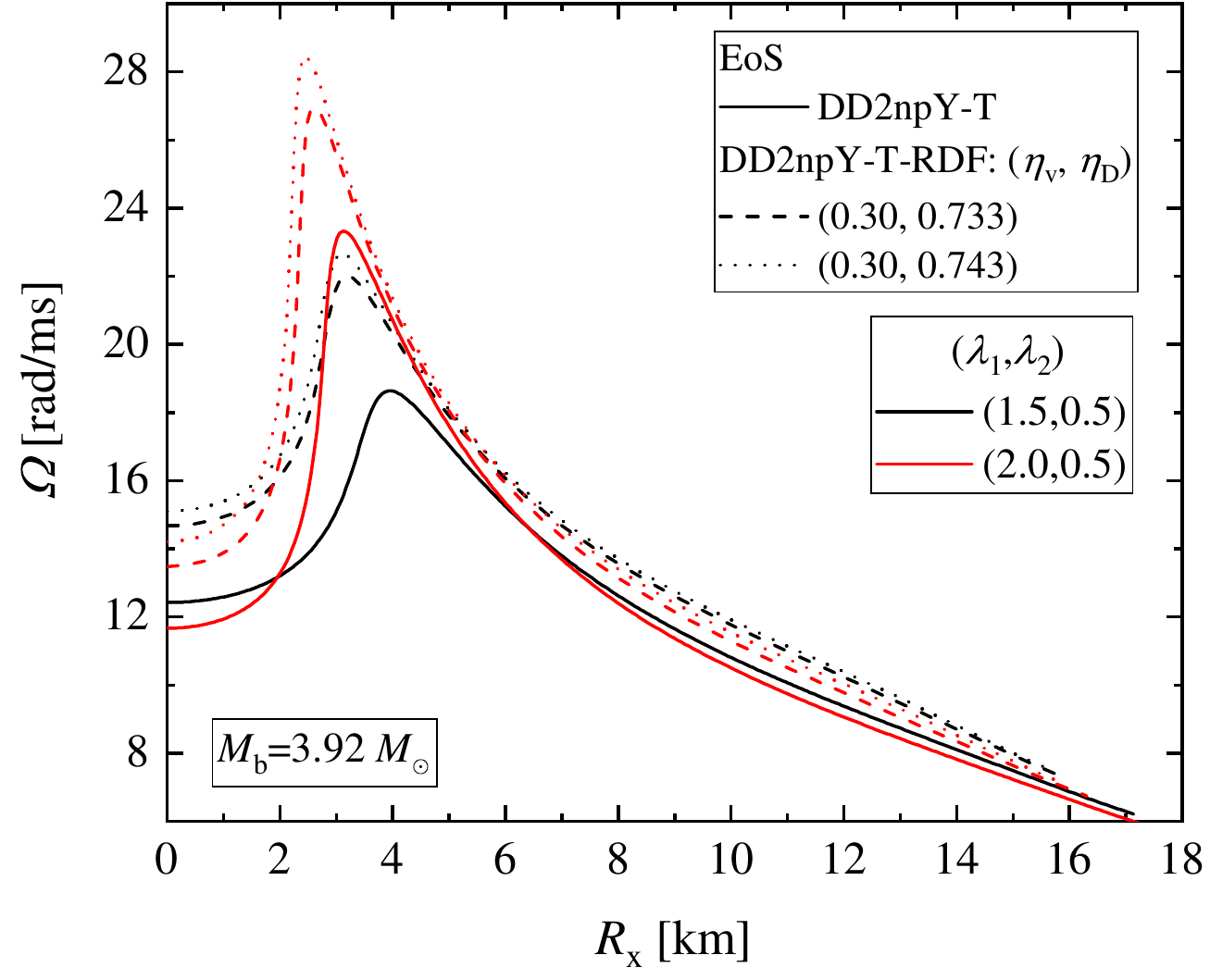}
	\caption{Angular velocity profiles for solutions with the same baryon mass $M_b=3.92 M_\odot$ and different combinations $(\lambda_1, \lambda_2)$ and EoS. Two solutions coincide with those shown in the bottom-right panel of Fig.~\ref{fig:rho_Om_r_tor}. }
    \label{fig:Om_R_tor_many}
\end{figure}

In Fig.~\ref{fig:rho_Om_r_tor} and Fig.~\ref{fig:special_tor} we have already studied and compared multiple different rotational profiles for quasi-toroidal solutions with different EoS both hybrid and hadronic. It is interesting, though, to compare the rotational profiles for a number of solutions with different EoS parameterizations and pairs $(\lambda_1,\lambda_2)$ on the same plot. In order to do that, in Fig.~\ref{fig:Om_R_tor_many} we compare the radial distribution of the angular velocity profiles for solutions with the hadronic and two hybrid EoSs and both pairs $(\lambda_1,\lambda_2)$ used in the quasi-toroidal case. Two of those solutions we already studied in the bottom right panel of Fig.~\ref{fig:rho_Om_r_tor}. Between the two $(\lambda_1,\lambda_2)$ cases, the solutions with higher $\lambda_1$ has steeper and higher peak in the angular velocity which is in corelation with the definition of $\lambda_1$. As we can already expect, no signature of the phase transition or  quark matter, being quark core or toroidal structure, can be observed in the rotational profile, and no major differences are induced due to the change in the EoS or $(\lambda_1,\lambda_2)$.  

\begin{figure}[ht]
\includegraphics[width=0.45\textwidth]{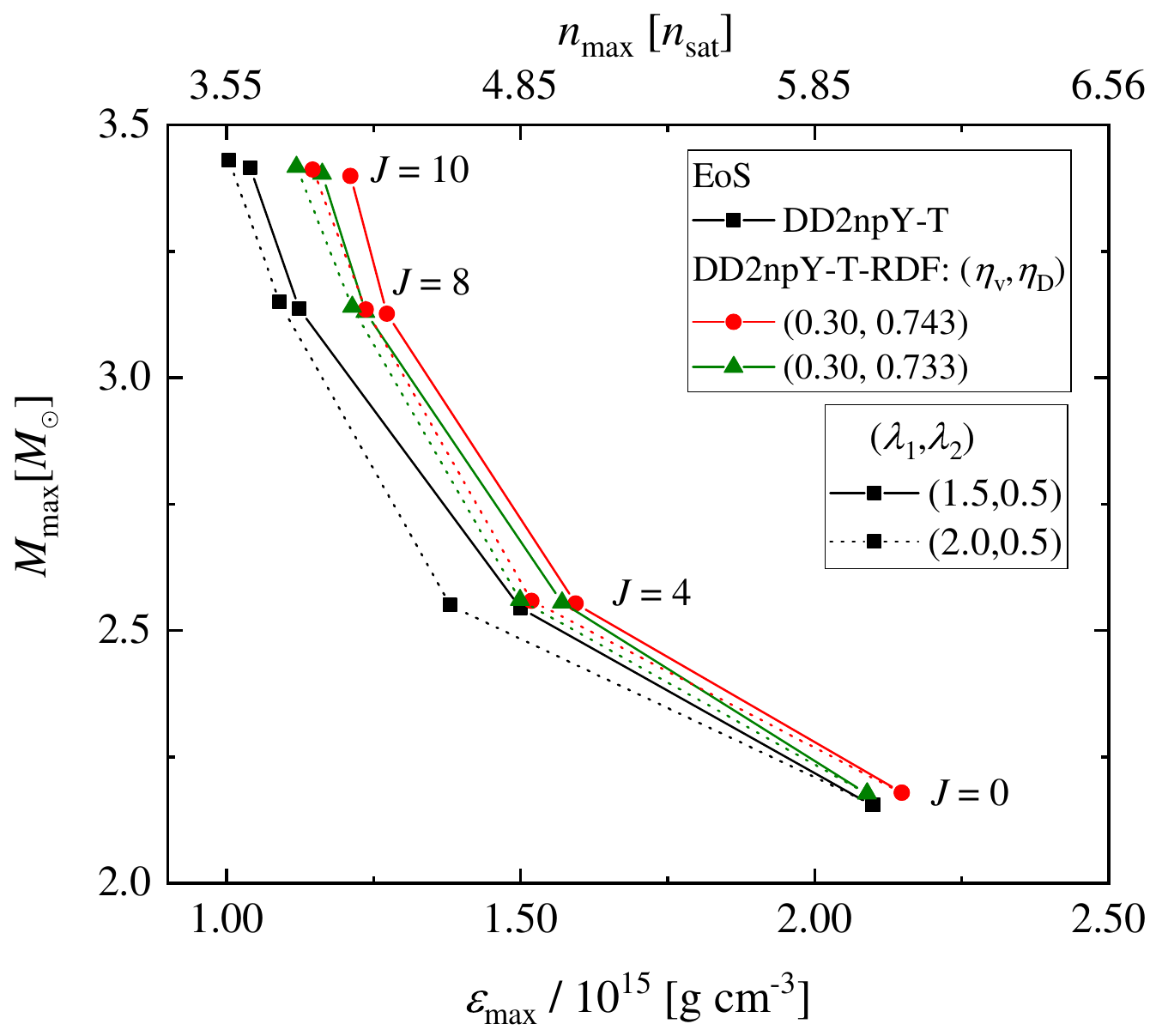}
	\caption{The maximum mass as a function of the maximum energy density for two hybrid and the hadronic EoSs. Solutions for both $(\lambda_1,\lambda_2) = (1.5,0.5)$ (those shown in Fig.~\ref{fig:Mmax_tor}) and for $(\lambda_1,\lambda_2) = (2.0,0.5)$ are compared. The change in $\lambda_1$ does not affect the behavior of the maximum mass models significantly.}
    \label{fig:Mmax_tor_2}
\end{figure}

In Fig.~\ref{fig:Mmax_tor_2} we extend the study of the shift of the maximum mass presented in Fig.~\ref{fig:Mmax_tor} for the same combinations of parameters and EoS parameterizations presented in Fig.~\ref{fig:Om_R_tor_many}. Changes in parameters $(\lambda_1,\lambda_2)$ have no significant impact on the values or the overall behavior of the maximum mass for the rotation rates considered.

\begin{figure}[ht]
    \includegraphics[width=0.45\textwidth]{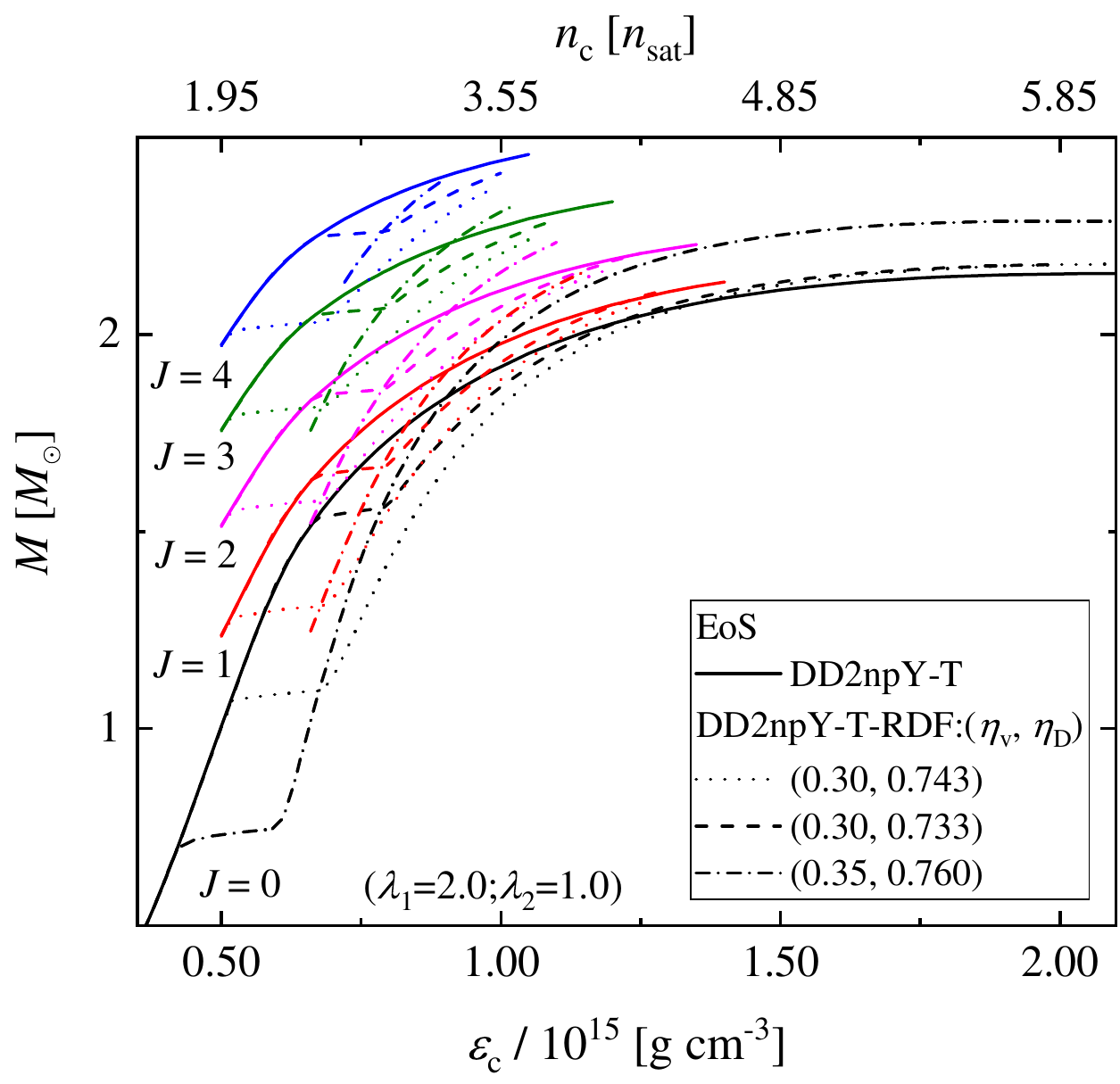}
    \includegraphics[width=0.45\textwidth]{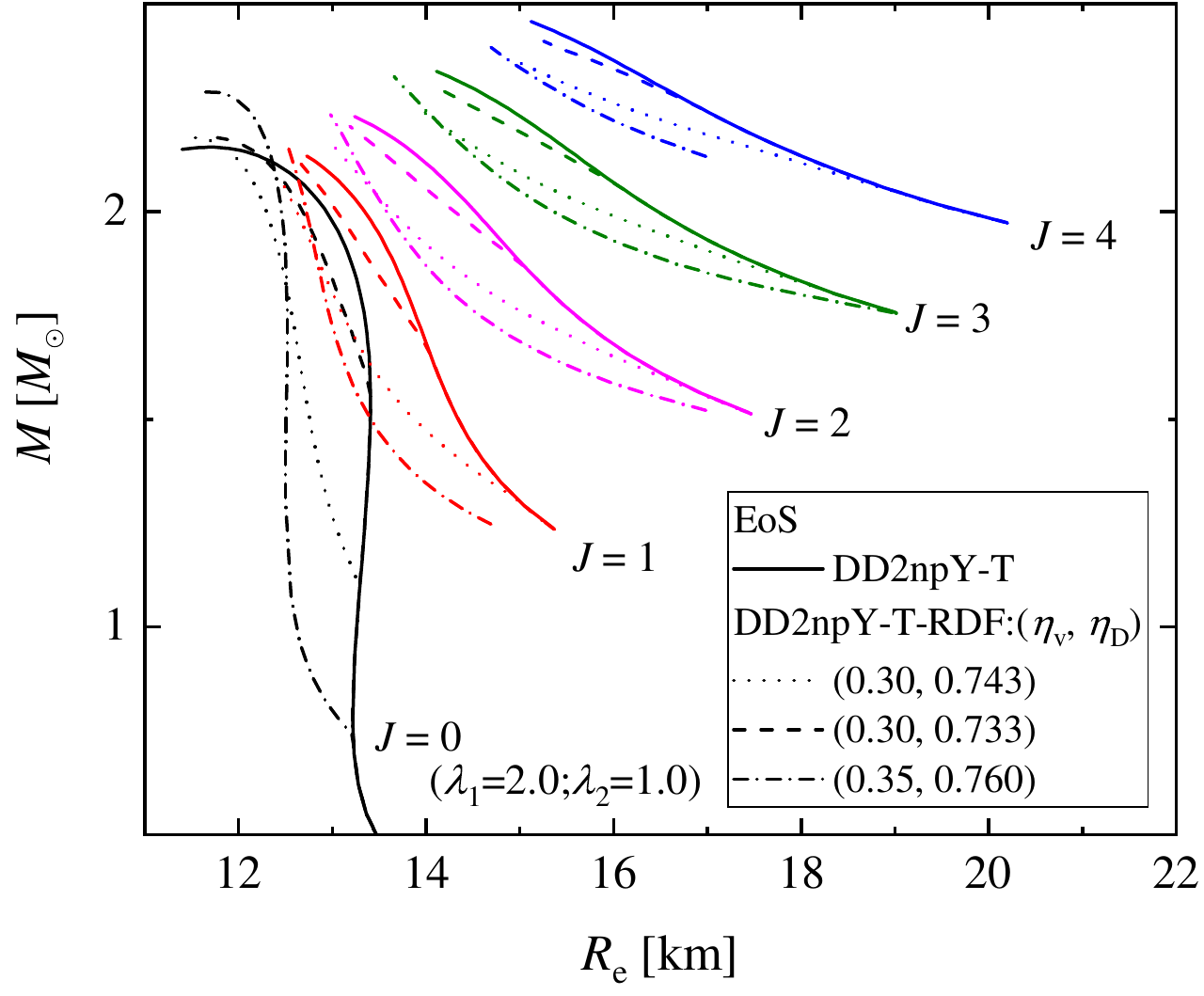}
	\caption{Quasi-spherical solutions with $(\lambda_1, \lambda_2) = (2.0,1.0)$. All of the notations and the EoS are the same as Fig. \ref{fig:M_rho_sph}.}
    \label{fig:M_rho_sph_2}
\end{figure}

In Fig.~\ref{fig:M_rho_sph_2} we proceed with the second pair $(\lambda_1,\lambda_2) = (2.0,1.0)$ for the quasi-spherical solutions. The higher value of $\lambda_1$ allows us to reach higher rotation rates than those shown in Fig.~\ref{fig:M_rho_sph}. However, the corresponding solutions are located at the low energy density end of the branches, resulting in progressively narrower energy density intervals over which hybrid star solutions exist. These increasingly short hybrid branches provide limited additional insight; therefore, for consistency with Fig.~\ref{fig:M_rho_sph}, we restrict the presented results to configurations with $J \leq 4$.
\bibliography{bibliography}

@article{Baiotti:2016qnr,
    author = "Baiotti, Luca and Rezzolla, Luciano",
    title = "{Binary neutron star mergers: a review of Einstein's richest laboratory}",
    eprint = "1607.03540",
    archivePrefix = "arXiv",
    primaryClass = "gr-qc",
    doi = "10.1088/1361-6633/aa67bb",
    journal = "Rept. Prog. Phys.",
    volume = "80",
    number = "9",
    pages = "096901",
    year = "2017"
}

@article{Iosif:2020iho,
    author = "Iosif, Panagiotis and Stergioulas, Nikolaos",
    title = "{Equilibrium sequences of differentially rotating stars with post-merger-like rotational profiles}",
    eprint = "2011.10612",
    archivePrefix = "arXiv",
    primaryClass = "gr-qc",
    reportNumber = "VIR-1007A-20",
    doi = "10.1093/mnras/stab392",
    journal = "Mon. Not. Roy. Astron. Soc.",
    volume = "503",
    number = "1",
    pages = "850--866",
    year = "2021"
}

@article{Ciolfi2021,
author={Ciolfi, Riccardo
and Stratta, Giulia
and Branchesi, Marica
and Gendre, Bruce},
title={Multi-messenger astrophysics with THESEUS in the 2030s},
journal={Experimental Astronomy},
year={2021},
month={Dec},
day={01},
volume={52},
number={3},
pages={245-275},
issn={1572-9508},
doi={10.1007/s10686-021-09795-9},
url={https://doi.org/10.1007/s10686-021-09795-9}
}

@article{Rosati2021,
author={Rosati, P.
and Basa, S.
and Blain, A. W.
and Bozzo, E.
and Branchesi, M.},
title={Synergies of THESEUS with the large facilities of the 2030s and guest observer opportunities},
journal={Experimental Astronomy},
year={2021},
month={Dec},
day={01},
volume={52},
number={3},
pages={407-437},
issn={1572-9508},
doi={10.1007/s10686-021-09764-2},
url={https://doi.org/10.1007/s10686-021-09764-2}
}

@article{Chakrabarty:2003kt,
    author = "Chakrabarty, Deepto and Morgan, Edward H. and Muno, Michael P. and Galloway, Duncan K. and Wijnands, Rudy and van der Klis, Michiel and Markwardt, Craig B.",
    title = "{Nuclear-powered millisecond pulsars and the maximum spin frequency of neutron stars}",
    eprint = "astro-ph/0307029",
    archivePrefix = "arXiv",
    doi = "10.1038/nature01732",
    journal = "Nature",
    volume = "424",
    pages = "42--44",
    year = "2003"
}

@article{Haskell:2018nlh,
    author = "Haskell, B. and Zdunik, J. L. and Fortin, M. and Bejger, M. and Wijnands, R. and Patruno, A.",
    title = "{Fundamental physics and the absence of sub-millisecond pulsars}",
    eprint = "1805.11277",
    archivePrefix = "arXiv",
    primaryClass = "astro-ph.HE",
    doi = "10.1051/0004-6361/201833521",
    journal = "Astron. Astrophys.",
    volume = "620",
    pages = "A69",
    year = "2018"
}

@article{Bozzola:2017qbu,
    author = "Bozzola, Gabriele and Stergioulas, Nikolaos and Bauswein, Andreas",
    title = "{Universal relations for differentially rotating relativistic stars at the threshold to collapse}",
    eprint = "1709.02787",
    archivePrefix = "arXiv",
    primaryClass = "gr-qc",
    doi = "10.1093/mnras/stx3002",
    journal = "Mon. Not. Roy. Astron. Soc.",
    volume = "474",
    number = "3",
    pages = "3557--3564",
    year = "2018"
}

@article{Ozel:2016oaf,
    author = {{\"O}zel, Feryal and Freire, Paulo},
    title = "{Masses, Radii, and the Equation of State of Neutron Stars}",
    eprint = "1603.02698",
    archivePrefix = "arXiv",
    primaryClass = "astro-ph.HE",
    doi = "10.1146/annurev-astro-081915-023322",
    journal = "Ann. Rev. Astron. Astrophys.",
    volume = "54",
    pages = "401--440",
    year = "2016"
}

@article{Paschalidis:2016vmz,
    author = "Paschalidis, Vasileios and Stergioulas, Nikolaos",
    title = "{Rotating Stars in Relativity}",
    eprint = "1612.03050",
    archivePrefix = "arXiv",
    primaryClass = "astro-ph.HE",
    doi = "10.1007/s41114-017-0008-x",
    journal = "Living Rev. Rel.",
    volume = "20",
    number = "1",
    pages = "7",
    year = "2017"
}

@article{Camelio:2020mdi,
    author = "Camelio, Giovanni and Dietrich, Tim and Rosswog, Stephan and Haskell, Brynmor",
    title = "{Axisymmetric models for neutron star merger remnants with realistic thermal and rotational profiles}",
    eprint = "2011.10557",
    archivePrefix = "arXiv",
    primaryClass = "astro-ph.HE",
    doi = "10.1103/PhysRevD.103.063014",
    journal = "Phys. Rev. D",
    volume = "103",
    number = "6",
    pages = "063014",
    year = "2021"
}

@article{Camelio:2019rsz,
    author = "Camelio, Giovanni and Dietrich, Tim and Marques, Miguel and Rosswog, Stephan",
    title = "{Rotating neutron stars with nonbarotropic thermal profile}",
    eprint = "1908.11258",
    archivePrefix = "arXiv",
    primaryClass = "gr-qc",
    doi = "10.1103/PhysRevD.100.123001",
    journal = "Phys. Rev. D",
    volume = "100",
    number = "12",
    pages = "123001",
    year = "2019"
}

@article{Iosif:2021aum,
    author = "Iosif, Panagiotis and Stergioulas, Nikolaos",
    title = "{Models of binary neutron star remnants with tabulated equations of state}",
    eprint = "2104.13672",
    archivePrefix = "arXiv",
    primaryClass = "astro-ph.HE",
    reportNumber = "VIR-0359B-21",
    doi = "10.1093/mnras/stab3565",
    journal = "Mon. Not. Roy. Astron. Soc.",
    volume = "510",
    number = "2",
    pages = "2948--2967",
    year = "2022",
    note = "[Erratum: Mon.Not.Roy.Astron.Soc. 517, 1601 (2022)]"
}

@article{Andersson:2000mf,
    author = "Andersson, Nils and Kokkotas, Kostas D.",
    title = "{The R mode instability in rotating neutron stars}",
    eprint = "gr-qc/0010102",
    archivePrefix = "arXiv",
    doi = "10.1142/S0218271801001062",
    journal = "Int. J. Mod. Phys. D",
    volume = "10",
    pages = "381--442",
    year = "2001"
}

@article{Hessels:2006ze,
    author = "Hessels, Jason W. T. and Ransom, Scott M. and Stairs, Ingrid H. and Freire, Paulo Cesar Carvalho and Kaspi, Victoria M. and Camilo, Fernando",
    title = "{A radio pulsar spinning at 716-hz}",
    eprint = "astro-ph/0601337",
    archivePrefix = "arXiv",
    doi = "10.1126/science.1123430",
    journal = "Science",
    volume = "311",
    pages = "1901--1904",
    year = "2006"
}

@article{Baym:2017whm,
    author = "Baym, Gordon and Hatsuda, Tetsuo and Kojo, Toru and Powell, Philip D. and Song, Yifan and Takatsuka, Tatsuyuki",
    title = "{From hadrons to quarks in neutron stars: a review}",
    eprint = "1707.04966",
    archivePrefix = "arXiv",
    primaryClass = "astro-ph.HE",
    reportNumber = "RIKEN-ITHEMS-REPORT-17, RIKEN-QHP-316, RIKEN-iTHEMS-Report-17",
    doi = "10.1088/1361-6633/aaae14",
    journal = "Rept. Prog. Phys.",
    volume = "81",
    number = "5",
    pages = "056902",
    year = "2018"
}

@article{Cipriani:2025wem,
    author = "Cipriani, Lorenzo and Sagun, Violetta and Staykov, Kalin V. and Doneva, Daniela D. and Yazadjiev, Stoytcho S.",
    title = "{Differentially rotating neutron stars with dark matter cores}",
    eprint = "2512.05898",
    archivePrefix = "arXiv",
    primaryClass = "astro-ph.HE",
    month = "12",
    year = "2025"
}

@article{Cierniak:2020eyh,
    author = "Cierniak, Mateusz and Blaschke, David",
    title = "{The special point on the hybrid star mass{\textendash}radius diagram and its multi{\textendash}messenger implications}",
    eprint = "2009.12353",
    archivePrefix = "arXiv",
    primaryClass = "astro-ph.HE",
    doi = "10.1140/epjst/e2020-000235-5",
    journal = "Eur. Phys. J. ST",
    volume = "229",
    number = "22-23",
    pages = "3663--3673",
    year = "2020"
}

@article{Cierniak:2021knt,
    author = "Cierniak, Mateusz and Blaschke, David",
    title = "{Hybrid neutron stars in the mass-radius diagram}",
    eprint = "2106.06986",
    archivePrefix = "arXiv",
    primaryClass = "nucl-th",
    doi = "10.1002/asna.202114000",
    journal = "Astron. Nachr.",
    volume = "342",
    number = "5",
    pages = "819--825",
    year = "2021"
}

@article{Cierniak:2021vlf,
    author = "Cierniak, Mateusz and Blaschke, David",
    title = "{Locating the special point of hybrid neutron stars}",
    eprint = "2112.09166",
    archivePrefix = "arXiv",
    primaryClass = "nucl-th",
    doi = "10.1051/epjconf/202225807009",
    journal = "EPJ Web Conf.",
    volume = "258",
    pages = "07009",
    year = "2022"
}

@article{Gill:2019bvq,
    author = "Gill, Ramandeep and Nathanail, Antonios and Rezzolla, Luciano",
    title = "{When Did the Remnant of GW170817 Collapse to a Black Hole?}",
    eprint = "1901.04138",
    archivePrefix = "arXiv",
    primaryClass = "astro-ph.HE",
    doi = "10.3847/1538-4357/ab16da",
    journal = "Astrophys. J.",
    volume = "876",
    number = "2",
    pages = "139",
    year = "2019"
}

@article{Cassing:2024dxp,
    author = "Cassing, Marie and Rezzolla, Luciano",
    title = "{Realistic models of general-relativistic differentially rotating stars}",
    eprint = "2405.06609",
    archivePrefix = "arXiv",
    primaryClass = "gr-qc",
    doi = "10.1093/mnras/stae1527",
    journal = "Mon. Not. Roy. Astron. Soc.",
    volume = "532",
    number = "1",
    pages = "945--964",
    year = "2024"
}

@article{Komatsu:1989ikr,
    author = "Komatsu, Hidemi and Eriguchi, Yoshiharu and Hachisu, Izumi",
    title = "{Rapidly rotating general relativistic stars {\textendash} I. Numerical method and its application to uniformly rotating polytropes}",
    doi = "10.1093/mnras/237.2.355",
    journal = "Mon. Not. Roy. Astron. Soc.",
    volume = "237",
    number = "2",
    pages = "355--379",
    year = "1989"
}

@article{Hammond:2025kki,
    author = "Hammond, P. and others",
    title = "{Investigating the Impact of Higher-Order Phase Transitions in Binary Neutron-Star Mergers}",
    eprint = "2508.10698",
    archivePrefix = "arXiv",
    primaryClass = "astro-ph.HE",
    month = "8",
    year = "2025"
}

@article{Han:2018mtj,
    author = "Han, Sophia and Steiner, Andrew W.",
    title = "{Tidal deformability with sharp phase transitions in (binary) neutron stars}",
    eprint = "1810.10967",
    archivePrefix = "arXiv",
    primaryClass = "nucl-th",
    doi = "10.1103/PhysRevD.99.083014",
    journal = "Phys. Rev. D",
    volume = "99",
    number = "8",
    pages = "083014",
    year = "2019"
}

@article{Ujevic:2022nkr,
    author = "Ujevic, Maximiliano and Gieg, Henrique and Schianchi, Federico and Chaurasia, Swami Vivekanandji and Tews, Ingo and Dietrich, Tim",
    title = "{Reverse phase transitions in binary neutron-star systems with exotic-matter cores}",
    eprint = "2211.04662",
    archivePrefix = "arXiv",
    primaryClass = "gr-qc",
    reportNumber = "Report-no: LA-UR-22-31740",
    doi = "10.1103/PhysRevD.107.024025",
    journal = "Phys. Rev. D",
    volume = "107",
    number = "2",
    pages = "024025",
    year = "2023"
}

@article{Blacker:2024tet,
    author = "Blacker, Sebastian and Bauswein, Andreas",
    title = "{Comprehensive survey of hybrid equations of state in neutron star mergers and constraints on the hadron-quark phase transition}",
    eprint = "2406.14669",
    archivePrefix = "arXiv",
    primaryClass = "astro-ph.HE",
    doi = "10.1103/31yz-4vnv",
    journal = "Phys. Rev. D",
    volume = "112",
    number = "12",
    pages = "123041",
    year = "2025"
}

@article{Chamel:2012ea,
    author = "Chamel, N. and Fantina, A. F. and Pearson, J. M. and Goriely, S.",
    title = "{Maximum mass of neutron stars with exotic cores}",
    eprint = "1205.0983",
    archivePrefix = "arXiv",
    primaryClass = "nucl-th",
    doi = "10.1051/0004-6361/201220986",
    journal = "Astron. Astrophys.",
    volume = "553",
    pages = "A22",
    year = "2013"
}

@article{Gartlein:2025zhd,
    author = {G{\"a}rtlein, Christoph and Ivanytskyi, Oleksii and Sagun, Violetta and Lopes, Il{\'\i}dio},
    title = "{Color-superconducting quarkyonic matter}",
    eprint = "2509.03517",
    archivePrefix = "arXiv",
    primaryClass = "nucl-th",
    month = "9",
    year = "2025"
}

@article{Ivanytskyi:2019ojt,
    author = "Ivanytskyi, O. and P{\'e}rez-Garc{\'\i}a, M. {\'A}ngeles and Sagun, V. and Albertus, C.",
    title = "{Second look to the Polyakov loop Nambu{\textendash}Jona-Lasinio model at finite baryonic density}",
    eprint = "1909.07421",
    archivePrefix = "arXiv",
    primaryClass = "hep-ph",
    doi = "10.1103/PhysRevD.100.103020",
    journal = "Phys. Rev. D",
    volume = "100",
    number = "10",
    pages = "103020",
    year = "2019"
}

@article{Somasundaram:2021clp,
    author = "Somasundaram, Rahul and Tews, Ingo and Margueron, J{\'e}r{\^o}me",
    title = "{Investigating signatures of phase transitions in neutron-star cores}",
    eprint = "2112.08157",
    archivePrefix = "arXiv",
    primaryClass = "nucl-th",
    reportNumber = "LA-UR-21-22340",
    doi = "10.1103/PhysRevC.107.025801",
    journal = "Phys. Rev. C",
    volume = "107",
    number = "2",
    pages = "025801",
    year = "2023"
}

@article{Jaisawal:2024lps,
    author = "Jaisawal, Gaurava K. and others",
    title = "{A Comprehensive Study of Thermonuclear X-ray Bursts from 4U 1820-30 with NICER: Accretion Disk Interactions and a Candidate Burst Oscillation}",
    eprint = "2504.07328",
    archivePrefix = "arXiv",
    primaryClass = "astro-ph.HE",
    doi = "10.3847/1538-4357/ad794e",
    journal = "Astrophys. J.",
    volume = "975",
    pages = "67",
    year = "2024"
}

@Article{Iosif:2021-09312,
    AUTHOR = {Iosif, Panagiotis and Stergioulas, Nikolaos},
    TITLE = {Differentially Rotating Relativistic Stars beyond the J-Constant Law},
    JOURNAL = {Physical Sciences Forum},
    VOLUME = {2},
    YEAR = {2021},
    NUMBER = {1},
    ARTICLE-NUMBER = {62},
    URL = {https://www.mdpi.com/2673-9984/2/1/62},
    ISSN = {2673-9984},
    DOI = {10.3390/ECU2021-09312}
}

@article{Oppenheimer:1939ne,
  title = {On Massive Neutron Cores},
  author = {Oppenheimer, J. R. and Volkoff, G. M.},
  journal = {Phys. Rev.},
  volume = {55},
  issue = {4},
  pages = {374--381},
  numpages = {0},
  year = {1939},
  month = {Feb},
  publisher = {American Physical Society},
  doi = {10.1103/PhysRev.55.374},
  url = {https://link.aps.org/doi/10.1103/PhysRev.55.374}
}

@article{Passamonti:2020yvh,
    author = "Passamonti, A. and Andersson, N.",
    title = "{Merger-inspired rotation laws and the low-T/W instability in neutron stars}",
    eprint = "2003.10198",
    archivePrefix = "arXiv",
    primaryClass = "astro-ph.SR",
    doi = "10.1093/mnras/staa2725",
    journal = "Mon. Not. Roy. Astron. Soc.",
    volume = "498",
    number = "4",
    pages = "5904--5915",
    year = "2020"
}

@article{Jaraba:2026nmq,
    author = "Jaraba, Santiago and Novak, J{\'e}r{\^o}me and Oertel, Micaela",
    title = "{Numerical simulations of oscillating and differentially rotating neutron stars}",
    eprint = "2601.10550",
    archivePrefix = "arXiv",
    primaryClass = "gr-qc",
    month = "1",
    year = "2026"
}

@article{Alford:2006vz,
    author = "Alford, Mark and Blaschke, David and Drago, Alessandro and Klahn, Thomas and Pagliara, Giuseppe and Schaffner-Bielich, Juergen",
    title = "{Quark matter in compact stars?}",
    eprint = "astro-ph/0606524",
    archivePrefix = "arXiv",
    doi = "10.1038/nature05582",
    journal = "Nature",
    volume = "445",
    pages = "E7--E8",
    year = "2007"
}

@article{Christian:2023hez,
    author = {Christian, Jan-Erik and Schaffner-Bielich, J{\"u}rgen and Rosswog, Stephan},
    title = "{Which first order phase transitions to quark matter are possible in neutron stars?}",
    eprint = "2312.10148",
    archivePrefix = "arXiv",
    primaryClass = "nucl-th",
    doi = "10.1103/PhysRevD.109.063035",
    journal = "Phys. Rev. D",
    volume = "109",
    number = "6",
    pages = "063035",
    year = "2024"
}

@article{Bauswein:2018bma,
    author = "Bauswein, Andreas and Bastian, Niels-Uwe F. and Blaschke, David B. and Chatziioannou, Katerina and Clark, James A. and Fischer, Tobias and Oertel, Micaela",
    title = "{Identifying a first-order phase transition in neutron star mergers through gravitational waves}",
    eprint = "1809.01116",
    archivePrefix = "arXiv",
    primaryClass = "astro-ph.HE",
    doi = "10.1103/PhysRevLett.122.061102",
    journal = "Phys. Rev. Lett.",
    volume = "122",
    number = "6",
    pages = "061102",
    year = "2019"
}

@article{Pili:2016hqo,
    author = "Pili, A. G. and Bucciantini, N. and Drago, A. and Pagliara, G. and Del Zanna, L.",
    title = "{Quark deconfinement in the proto-magnetar model of long gamma-ray bursts}",
    eprint = "1606.02075",
    archivePrefix = "arXiv",
    primaryClass = "astro-ph.HE",
    doi = "10.1093/mnrasl/slw115",
    journal = "Mon. Not. Roy. Astron. Soc.",
    volume = "462",
    number = "1",
    pages = "L26--L30",
    year = "2016"
}

@article{Tootle:2026ebk,
    author = "Tootle, Samuel D. and Jacques, Terrence Pierre and Cassing, Marie",
    title = "{A new code for computing differentially rotating neutron stars}",
    eprint = "2601.05176",
    archivePrefix = "arXiv",
    primaryClass = "gr-qc",
    month = "1",
    year = "2026"
}

@article{Franceschetti:2022ypc,
    author = "Franceschetti, Kevin and Del Zanna, Luca and Soldateschi, Jacopo and Bucciantini, Niccol{\`o}",
    title = "{Numerical Equilibrium Configurations and Quadrupole Moments of Post-Merger Differentially Rotating Relativistic Stars}",
    doi = "10.3390/universe8030172",
    journal = "Universe",
    volume = "8",
    number = "3",
    pages = "172",
    year = "2022"
}

@misc{lorene,
  title        = {LORENE: Langage Objet pour la Relativit{\'e} Num{\'e}rique},
  howpublished = {\url{http://www.lorene.obspm.fr}},
  note         = {Numerical relativity library}
}

@article{Weih:2019xvw,
    author = "Weih, Lukas R. and Hanauske, Matthias and Rezzolla, Luciano",
    title = "{Postmerger Gravitational-Wave Signatures of Phase Transitions in Binary Mergers}",
    eprint = "1912.09340",
    archivePrefix = "arXiv",
    primaryClass = "gr-qc",
    doi = "10.1103/PhysRevLett.124.171103",
    journal = "Phys. Rev. Lett.",
    volume = "124",
    number = "17",
    pages = "171103",
    year = "2020"
}

@article{Zhang:2017fsy,
    author = "Zhang, Xuefeng and Cao, Zhoujian and Gao, He",
    title = "{Long-term postmerger simulations of relativistic star coalescence: Formation of toroidal remnants and gravitational wave afterglow}",
    eprint = "1710.01881",
    archivePrefix = "arXiv",
    primaryClass = "gr-qc",
    doi = "10.1142/S0218271819500263",
    journal = "Int. J. Mod. Phys. D",
    volume = "28",
    number = "01",
    pages = "1950026",
    year = "2018"
}

@article{Paschalidis:2015mla,
    author = "Paschalidis, Vasileios and East, William E. and Pretorius, Frans and Shapiro, Stuart L.",
    title = "{One-arm Spiral Instability in Hypermassive Neutron Stars Formed by Dynamical-Capture Binary Neutron Star Mergers}",
    eprint = "1510.03432",
    archivePrefix = "arXiv",
    primaryClass = "astro-ph.HE",
    doi = "10.1103/PhysRevD.92.121502",
    journal = "Phys. Rev. D",
    volume = "92",
    number = "12",
    pages = "121502",
    year = "2015"
}

@article{Most:2018eaw,
    author = {Most, Elias R. and Papenfort, L. Jens and Dexheimer, Veronica and Hanauske, Matthias and Schramm, Stefan and St{\"o}cker, Horst and Rezzolla, Luciano},
    title = "{Signatures of quark-hadron phase transitions in general-relativistic neutron-star mergers}",
    eprint = "1807.03684",
    archivePrefix = "arXiv",
    primaryClass = "astro-ph.HE",
    doi = "10.1103/PhysRevLett.122.061101",
    journal = "Phys. Rev. Lett.",
    volume = "122",
    number = "6",
    pages = "061101",
    year = "2019"
}

@article{Prakash:2021wpz,
    author = "Prakash, Aviral and Radice, David and Logoteta, Domenico and Perego, Albino and Nedora, Vsevolod and Bombaci, Ignazio and Kashyap, Rahul and Bernuzzi, Sebastiano and Endrizzi, Andrea",
    title = "{Signatures of deconfined quark phases in binary neutron star mergers}",
    eprint = "2106.07885",
    archivePrefix = "arXiv",
    primaryClass = "astro-ph.HE",
    doi = "10.1103/PhysRevD.104.083029",
    journal = "Phys. Rev. D",
    volume = "104",
    number = "8",
    pages = "083029",
    year = "2021"
}

@article{Bozzola:2019tit,
    author = "Bozzola, Gabriele and Espino, Pedro L. and Lewin, Collin D. and Paschalidis, Vasileios",
    title = "{Maximum mass and universal relations of rotating relativistic hybrid hadron-quark stars}",
    eprint = "1905.00028",
    archivePrefix = "arXiv",
    primaryClass = "astro-ph.HE",
    doi = "10.1140/epja/i2019-12831-2",
    journal = "Eur. Phys. J. A",
    volume = "55",
    number = "9",
    pages = "149",
    year = "2019"
}

@article{Duez:2004nf,
    author = "Duez, Matthew D. and Liu, Yuk Tung and Shapiro, Stuart L. and Stephens, Branson C.",
    title = "{General relativistic hydrodynamics with viscosity: Contraction, catastrophic collapse, and disk formation in hypermassive neutron stars}",
    eprint = "astro-ph/0402502",
    archivePrefix = "arXiv",
    doi = "10.1103/PhysRevD.69.104030",
    journal = "Phys. Rev. D",
    volume = "69",
    pages = "104030",
    year = "2004"
}

@article{Weissenborn:2011qu,
    author = {Weissenborn, Simon and Sagert, Irina and Pagliara, Giuseppe and Hempel, Matthias and Schaffner-Bielich, J{\"u}rgen},
    title = "{Quark Matter In Massive Neutron Stars}",
    eprint = "1102.2869",
    archivePrefix = "arXiv",
    primaryClass = "astro-ph.HE",
    doi = "10.1088/2041-8205/740/1/L14",
    journal = "Astrophys. J. Lett.",
    volume = "740",
    pages = "L14",
    year = "2011"
}

@article{Tolman:1939jz,
  title = {Static Solutions of Einstein's Field Equations for Spheres of Fluid},
  author = {Tolman, Richard C.},
  journal = {Phys. Rev.},
  volume = {55},
  issue = {4},
  pages = {364--373},
  numpages = {0},
  year = {1939},
  month = {Feb},
  publisher = {American Physical Society},
  doi = {10.1103/PhysRev.55.364},
  url = {https://link.aps.org/doi/10.1103/PhysRev.55.364}
}

@article{Shahrbaf:2022upc,
    author = "Shahrbaf, M. and Blaschke, D. and Typel, S. and Farrar, G. R. and Alvarez-Castillo, D. E.",
    title = "{Sexaquark dilemma in neutron stars and its solution by quark deconfinement}",
    eprint = "2202.00652",
    archivePrefix = "arXiv",
    primaryClass = "nucl-th",
    doi = "10.1103/PhysRevD.105.103005",
    journal = "Phys. Rev. D",
    volume = "105",
    number = "10",
    pages = "103005",
    year = "2022"
}

@article{Typel:2018wmm,
    author = "Typel, S.",
    title = "{Equations of state for astrophysical simulations from generalized relativistic density functionals}",
    doi = "10.1088/1361-6471/aadea5",
    journal = "J. Phys. G",
    volume = "45",
    number = "11",
    pages = "114001",
    year = "2018"
}

@article{Romani:2021xmb,
    author = "Romani, Roger W. and Kandel, D. and Filippenko, Alexei V. and Brink, Thomas G. and Zheng, WeiKang",
    title = "{PSR J1810+1744: Companion Darkening and a Precise High Neutron Star Mass}",
    eprint = "2101.09822",
    archivePrefix = "arXiv",
    primaryClass = "astro-ph.HE",
    doi = "10.3847/2041-8213/abe2b4",
    journal = "Astrophys. J. Lett.",
    volume = "908",
    number = "2",
    pages = "L46",
    year = "2021"
}

@article{Vinciguerra:2023qxq,
    author = "Vinciguerra, Serena and others",
    title = "{An Updated Mass{\textendash}Radius Analysis of the 2017{\textendash}2018 NICER Data Set of PSR J0030+0451}",
    eprint = "2308.09469",
    archivePrefix = "arXiv",
    primaryClass = "astro-ph.HE",
    doi = "10.3847/1538-4357/acfb83",
    journal = "Astrophys. J.",
    volume = "961",
    number = "1",
    pages = "62",
    year = "2024"
}

@article{Salmi:2024aum,
    author = "Salmi, Tuomo and others",
    title = "{The Radius of the High-mass Pulsar PSR J0740+6620 with 3.6 yr of NICER Data}",
    eprint = "2406.14466",
    archivePrefix = "arXiv",
    primaryClass = "astro-ph.HE",
    doi = "10.3847/1538-4357/ad5f1f",
    journal = "Astrophys. J.",
    volume = "974",
    number = "2",
    pages = "294",
    year = "2024"
}

@article{Mauviard:2025dmd,
    author = "Mauviard, Lucien and others",
    title = "{A NICER View of the 1.4 M$_{⊙}$ Edge-on Pulsar PSR J0614-3329}",
    eprint = "2506.14883",
    archivePrefix = "arXiv",
    primaryClass = "astro-ph.HE",
    doi = "10.3847/1538-4357/ae145d",
    journal = "Astrophys. J.",
    volume = "995",
    number = "1",
    pages = "60",
    year = "2025"
}

@article{Ivanytskyi:2022bjc,
    author = "Ivanytskyi, Oleksii and Blaschke, David B.",
    title = "{Recovering the Conformal Limit of Color Superconducting Quark Matter within a Confining Density Functional Approach}",
    eprint = "2209.02050",
    archivePrefix = "arXiv",
    primaryClass = "nucl-th",
    doi = "10.3390/particles5040038",
    journal = "Particles",
    volume = "5",
    number = "4",
    pages = "514--534",
    year = "2022"
}

@article{Gartlein:2025,
    author = "Gärtlein, Christoph  and Sagun, Violetta and Ivanytskyi, Oleksii and Blaschke, David and Lopes, Ilídio",
    title = "{Rapidly Spinning Massive Pulsars as an Indicator of Quark Deconfinement}",
    eprint = "2512.07977",
    archivePrefix = "arXiv",
    primaryClass = "nucl-th",
    doi = "10.48550/arXiv.2512.07977",
    year = "2025"
}

@article{Gartlein:2023vif,
    author = {G{\"a}rtlein, Christoph and Ivanytskyi, Oleksii and Sagun, Violetta and Blaschke, David},
    title = "{Hybrid star phenomenology from the properties of the special point}",
    eprint = "2301.10765",
    archivePrefix = "arXiv",
    primaryClass = "nucl-th",
    doi = "10.1103/PhysRevD.108.114028",
    journal = "Phys. Rev. D",
    volume = "108",
    number = "11",
    pages = "114028",
    year = "2023"
}

@article{Ivanytskyi:2022oxv,
    author = "Ivanytskyi, Oleksii and Blaschke, David",
    title = "{Density functional approach to quark matter with confinement and color superconductivity}",
    eprint = "2204.03611",
    archivePrefix = "arXiv",
    primaryClass = "nucl-th",
    doi = "10.1103/PhysRevD.105.114042",
    journal = "Phys. Rev. D",
    volume = "105",
    number = "11",
    pages = "114042",
    year = "2022"
}

@article{Choudhury:2024xbk,
    author = "Choudhury, Devarshi and others",
    title = "{A NICER View of the Nearest and Brightest Millisecond Pulsar: PSR J0437\textendash{}4715}",
    eprint = "2407.06789",
    archivePrefix = "arXiv",
    primaryClass = "astro-ph.HE",
    doi = "10.3847/2041-8213/ad5a6f",
    journal = "Astrophys. J. Lett.",
    volume = "971",
    number = "1",
    pages = "L20",
    year = "2024"
}

@article{Salmi:2024bss,
    author = "Salmi, Tuomo and others",
    title = "{A NICER View of PSR J1231{\ensuremath{-}}1411: A Complex Case}",
    eprint = "2409.14923",
    archivePrefix = "arXiv",
    primaryClass = "astro-ph.HE",
    doi = "10.3847/1538-4357/ad81d2",
    journal = "Astrophys. J.",
    volume = "976",
    number = "1",
    pages = "58",
    year = "2024"
}

@article{Miller:2019cac,
    author = "Miller, M. C. and others",
    title = "{PSR J0030+0451 Mass and Radius from $NICER$ Data and Implications for the Properties of Neutron Star Matter}",
    eprint = "1912.05705",
    archivePrefix = "arXiv",
    primaryClass = "astro-ph.HE",
    doi = "10.3847/2041-8213/ab50c5",
    journal = "Astrophys. J. Lett.",
    volume = "887",
    number = "1",
    pages = "L24",
    year = "2019"
}

@article{Dittmann:2024mbo,
    author = "Dittmann, Alexander J. and others",
    title = "{A More Precise Measurement of the Radius of PSR J0740+6620 Using Updated NICER Data}",
    eprint = "2406.14467",
    archivePrefix = "arXiv",
    primaryClass = "astro-ph.HE",
    doi = "10.3847/1538-4357/ad5f1e",
    journal = "Astrophys. J.",
    volume = "974",
    number = "2",
    pages = "295",
    year = "2024"
}

@article{Fonseca:2021wxt,
    author = "Fonseca, E. and others",
    title = "{Refined Mass and Geometric Measurements of the High-mass PSR J0740+6620}",
    eprint = "2104.00880",
    archivePrefix = "arXiv",
    primaryClass = "astro-ph.HE",
    doi = "10.3847/2041-8213/ac03b8",
    journal = "Astrophys. J. Lett.",
    volume = "915",
    number = "1",
    pages = "L12",
    year = "2021"
}

@article{LIGOScientific:2018cki,
    author = "Abbott, B. P. and others",
    collaboration = "LIGO Scientific, Virgo",
    title = "{GW170817: Measurements of neutron star radii and equation of state}",
    eprint = "1805.11581",
    archivePrefix = "arXiv",
    primaryClass = "gr-qc",
    reportNumber = "LIGO-P1800115",
    doi = "10.1103/PhysRevLett.121.161101",
    journal = "Phys. Rev. Lett.",
    volume = "121",
    number = "16",
    pages = "161101",
    year = "2018"
}

@article{Glendenning:1992vb,
    author = "Glendenning, Norman K.",
    title = "{First order phase transitions with more than one conserved charge: Consequences for neutron stars}",
    doi = "10.1103/PhysRevD.46.1274",
    journal = "Phys. Rev. D",
    volume = "46",
    pages = "1274--1287",
    year = "1992"
}

@article{Staykov:2023ose,
    author = "Staykov, Kalin V. and Doneva, Daniela D. and Heisenberg, Lavinia and Stergioulas, Nikolaos and Yazadjiev, Stoytcho S.",
    title = "{Differentially rotating scalarized neutron stars with realistic postmerger profiles}",
    eprint = "2303.07769",
    archivePrefix = "arXiv",
    primaryClass = "gr-qc",
    reportNumber = "ET-0053A-23",
    doi = "10.1103/PhysRevD.108.024058",
    journal = "Phys. Rev. D",
    volume = "108",
    number = "2",
    pages = "024058",
    year = "2023"
}

@article{Annala:2019puf,
    author = {Annala, Eemeli and Gorda, Tyler and Kurkela, Aleksi and N{\"a}ttil{\"a}, Joonas and Vuorinen, Aleksi},
    title = "{Evidence for quark-matter cores in massive neutron stars}",
    eprint = "1903.09121",
    archivePrefix = "arXiv",
    primaryClass = "astro-ph.HE",
    reportNumber = "CERN-TH-2019-031, HIP-2019-7/TH",
    doi = "10.1038/s41567-020-0914-9",
    journal = "Nature Phys.",
    volume = "16",
    number = "9",
    pages = "907--910",
    year = "2020"
}

@article{Ansorg:2008pk,
    author = "Ansorg, Marcus and Gondek-Rosinska, Dorota and Villain, Loic",
    title = "{The Continuous Parametric Transition of Spheroidal to Toroidal Differentially Rotating Stars in General Relativity}",
    eprint = "0812.3347",
    archivePrefix = "arXiv",
    primaryClass = "gr-qc",
    doi = "10.1111/j.1365-2966.2009.14904.x",
    journal = "Mon. Not. Roy. Astron. Soc.",
    volume = "396",
    pages = "2359",
    year = "2009"
}

@ARTICLE{1989MNRAS.239..153K,
       author = {{Komatsu}, Hidemi and {Eriguchi}, Yoshiharu and {Hachisu}, Izumi},
        title = "{Rapidly rotating general relativistic stars. II - Differentially rotating polytropes}",
      journal =  "Mon. Not. Roy. Astron. Soc.",
         year = "1989",
       volume = "239",
        pages = "153-171",
          doi = "10.1093/mnras/239.1.153"
}

@article{Gartlein:2024cbj,
    author = {G{\"a}rtlein, Christoph and Sagun, Violetta and Ivanytskyi, Oleksii and Blaschke, David and Lopes, Ilidio},
    title = "{Fastest spinning millisecond pulsars: Indicators for quark matter in neutron stars?}",
    eprint = "2412.07758",
    archivePrefix = "arXiv",
    primaryClass = "nucl-th",
    doi = "10.1103/PhysRevD.111.123021",
    journal = "Phys. Rev. D",
    volume = "111",
    number = "12",
    pages = "123021",
    year = "2025"
}

@ARTICLE{1994ApJ...422..227C,
       author = {{Cook}, Gregory B. and {Shapiro}, Stuart L. and {Teukolsky}, Saul A.},
        title = "{Rapidly Rotating Polytropes in General Relativity}",
      journal = {\apj},
         year = 1994,
        month = feb,
       volume = {422},
        pages = {227},
          doi = {10.1086/173721}
}

@ARTICLE{1995ApJ...444..306S,
       author = {{Stergioulas}, Nikolaos and {Friedman}, John L.},
        title = "{Comparing Models of Rapidly Rotating Relativistic Stars Constructed by Two Numerical Methods}",
      journal = {\apj},
         year = 1995,
        month = may,
       volume = {444},
        pages = {306},
          doi = {10.1086/175605},
archivePrefix = {arXiv},
       eprint = {astro-ph/9411032},
 primaryClass = {astro-ph},
       adsurl = {https://ui.adsabs.harvard.edu/abs/1995ApJ...444..306S}
}

@article{Uryu:2017obi,
    author = "Uryu, Koji and Tsokaros, Antonios and Baiotti, Luca and Galeazzi, Filippo and Taniguchi, Keisuke and Yoshida, Shin'ichirou",
    title = "{Modeling differential rotations of compact stars in equilibriums}",
    eprint = "1709.02643",
    archivePrefix = "arXiv",
    primaryClass = "astro-ph.HE",
    doi = "10.1103/PhysRevD.96.103011",
    journal = "Phys. Rev. D",
    volume = "96",
    number = "10",
    pages = "103011",
    year = "2017"
}

@article{Lam:2025jsk,
    author = "Lam, Alan Tsz-Lok and Staykov, Kalin V. and Kuan, Hao-Jui and Doneva, Daniela D. and Yazadjiev, Stoytcho S.",
    title = "{Axisymmetric stability of neutron stars as extreme rotators in massive scalar-tensor theory}",
    eprint = "2502.03973",
    archivePrefix = "arXiv",
    primaryClass = "gr-qc",
    doi = "10.1103/PhysRevD.111.104030",
    journal = "Phys. Rev. D",
    volume = "111",
    number = "10",
    pages = "104030",
    year = "2025"
}

@article{Weih:2017mcw,
    author = "Weih, Lukas R. and Most, Elias R. and Rezzolla, Luciano",
    title = "{On the stability and maximum mass of differentially rotating relativistic stars}",
    eprint = "1709.06058",
    archivePrefix = "arXiv",
    primaryClass = "gr-qc",
    doi = "10.1093/mnrasl/slx178",
    journal = "Mon. Not. Roy. Astron. Soc.",
    volume = "473",
    number = "1",
    pages = "L126--L130",
    year = "2018"
}

@article{Szewczyk:2025gtz,
    author = "Szewczyk, Pawe{\l} and Gondek-Rosi{\'n}ska, Dorota and Cerd{\'a}-Dur{\'a}n, Pablo",
    title = "{Dynamical Stability of Hypermassive Neutron Stars against Quasi-radial Perturbations}",
    eprint = "2507.05453",
    archivePrefix = "arXiv",
    primaryClass = "astro-ph.HE",
    doi = "10.3847/1538-4357/adee12",
    journal = "Astrophys. J.",
    volume = "990",
    number = "2",
    pages = "199",
    year = "2025"
}

@article{Zhou:2019hyy,
    author = "Zhou, Enping and Tsokaros, Antonios and Uryu, Koji and Xu, Renxin and Shibata, Masaru",
    title = "{Differentially rotating strange star in general relativity}",
    eprint = "1902.09361",
    archivePrefix = "arXiv",
    primaryClass = "astro-ph.HE",
    doi = "10.1103/PhysRevD.100.043015",
    journal = "Phys. Rev. D",
    volume = "100",
    number = "4",
    pages = "043015",
    year = "2019"
}

@article{Hanauske:2016gia,
    author = {Hanauske, Matthias and Takami, Kentaro and Bovard, Luke and Rezzolla, Luciano and Font, Jos{\'e} A. and Galeazzi, Filippo and St{\"o}cker, Horst},
    title = "{Rotational properties of hypermassive neutron stars from binary mergers}",
    eprint = "1611.07152",
    archivePrefix = "arXiv",
    primaryClass = "gr-qc",
    doi = "10.1103/PhysRevD.96.043004",
    journal = "Phys. Rev. D",
    volume = "96",
    number = "4",
    pages = "043004",
    year = "2017"
}

@article{DePietri:2019mti,
    author = {De Pietri, Roberto and Feo, Alessandra and Font, Jos{\'e} A. and L{\"o}ffler, Frank and Pasquali, Michele and Stergioulas, Nikolaos},
    title = "{Numerical-relativity simulations of long-lived remnants of binary neutron star mergers}",
    eprint = "1910.04036",
    archivePrefix = "arXiv",
    primaryClass = "gr-qc",
    doi = "10.1103/PhysRevD.101.064052",
    journal = "Phys. Rev. D",
    volume = "101",
    number = "6",
    pages = "064052",
    year = "2020"
}

@article{Xie:2020udh,
    author = "Xie, Xiaoyi and Hawke, Ian and Passamonti, Andrea and Andersson, Nils",
    title = "{Instabilities in neutron-star postmerger remnants}",
    eprint = "2005.13696",
    archivePrefix = "arXiv",
    primaryClass = "astro-ph.HE",
    doi = "10.1103/PhysRevD.102.044040",
    journal = "Phys. Rev. D",
    volume = "102",
    number = "4",
    pages = "044040",
    year = "2020"
}

\end{document}